\PassOptionsToPackage{round, authoryear}{natbib} 
\documentclass[final,5p,times,twocolumn,authoryear]{elsarticle}

\usepackage[utf8]{inputenc}

\usepackage{setspace}
\usepackage{amssymb,amsmath,amsfonts,mathtools} 
\usepackage{graphicx}

\usepackage{hhline}
\usepackage{ulem}

\usepackage{tabularx}
\usepackage{multirow}
\usepackage{multicol}
\usepackage{array}
\usepackage[dvipsnames]{xcolor}
\usepackage{xfrac}
\usepackage{subcaption}
\usepackage{wrapfig}
\usepackage{lscape}
\usepackage{rotating}
\usepackage{epstopdf}
\usepackage{pbox}

\usepackage{natbib}

\usepackage[font=small, labelsep=period]{caption}
\usepackage{comment}

\newcommand\inv[1]{#1\raisebox{1.15ex}{$\scriptscriptstyle-\!1$}}
\newcommand{\spectralimaging}{SI}
\newcommand{\remotesensing}{RS}
\newcommand\spectralimage[1]{#1SI}
\newcommand{\relativemeasure}{RA}
\newcommand{\dlm}{DLM}
\newcommand{\vca}{VCA}
\newcommand\authorcont[1]{\textbf{\MakeUppercase{#1}\,:\,}}

\begin{document}

\title{Single-target mineral detection with site-specific endmember extraction for survey points identification: A case study of Jaffna, Sri Lanka}

\thispagestyle{empty}
\journal{journal}
\date{}

\author[add1]{D.~Y.~L.~Ranasinghe\corref{cor1}} \ead{yasirur@sltc.ac.lk}
\author[add2]{H.~M.~H.~K.~Weerasooriya} \ead{kavingaweerasooriya@eng.pdn.ac.lk}
\author[add3,add2]{S.~Herath} \ead{sanjayah@umd.edu}
\author[add2]{H.~M.~V.~R.~Herath} \ead{vijitha@eng.pdn.ac.lk}
\author[add2]{G.~M.~R.~I.~Godaliyadda} \ead{roshan.godd@ee.pdn.ac.lk}
\author[add2]{M.~P.~B.~Ekanayake} \ead{mpb@ee.pdn.ac.lk}
\author[add4]{A.~Senaratne} \ead{atulas@sci.pdn.ac.lk}
\author[add1]{S.~L.~P.~Yasakethu} \ead{lasithy@sltc.ac.lk}

\cortext[cor1]{Corresponding author}
\address[add1]{School of Engineering, Sri Lanka Technological Campus, Padukka, Sri Lanka, 10500}
\address[add2]{Dept. of Electrical and Electronic Engineering, Faculty of Engineering, University of Peradeniya, Peradeniya, Sri Lanka, 20400}
\address[add3]{Dept. of Electrical and Computer Engineering, University of Maryland, College Park, MD, USA, 20740}
\address[add4]{Dept. of Geology, Faculty of Science, University of Peradeniya, Peradeniya, Sri Lanka, 20400}

\setlength{\parindent}{5mm}

\begin{abstract}
As field surveys used for manual lithological mapping are costly and time-consuming, digital lithological mapping (\dlm) that utilizes remotely sensed spectral imaging provides a viable and economical alternative. Generally, \dlm~ has been performed using spectral imaging with the use of laboratory-generated generic endmember signatures. However, the use of generic signatures is error-prone due to the presence of site-specific impurification processes. To that end, this paper proposes generating a single-target abundance mineral map for \dlm, where the generated map can further be used as a guide for the selection or avoidance of a field survey.  For that, a stochastic cancellation-based methodology was used to generate a site-specific endemic signature for the mineral in concern to reduce the inclusive nature otherwise present in \dlm. Here, single-target detection allows the generation of a more accurate site-specific signature for lithological mapping as opposed to multi-target detection. Furthermore, a soil pixel alignment strategy to visualize the relative purity level of the target mineral has been introduced in the proposed work. Then, for the method validation, mapping of limestone deposits in the Jaffna peninsula of Sri Lanka was conducted as the case study using satellite-based spectral imaging as the input. It was observed that despite the low signal-to-noise ratio of the input hyperspectral data the proposed methodology was able to robustly extract the rich information contained in the input data. Further, a field survey was conducted to collect soil samples of four sites chosen by the proposed \dlm~ from the Jaffna peninsula as an algorithm validation and to demonstrate the application of the proposed solution. The proposed abundance threshold of 0.1 -- a hard decision boundary for limestone presence and absence -- coincided with the industrial standard X-ray diffraction (XRD) threshold of 5\,\% for the mineral presence. The results of the XRD test validated the use of the algorithm in the selection of sites to be surveyed, hence could avoid conducting a costly field survey on the assumption of the existence of a mineral. Then a more rigorous survey may be performed, if required, only if the proposed digital lithological survey affirms. 
\end{abstract}


\maketitle

\section{Introduction}
\label{section: introduction}
Over the recent years, spectral imaging (\spectralimaging)  has been used to ascertain information regarding diverse fields, including the climate and environment \citep{henderson1997sar}, biodiversity \citep{randin2020monitoring}, eco-systems \citep{donlon2012global}, and food quality \citep{bandara2020validation, weerasooriya2020transmittance}, etc. The non-intrusive nature and the capacity to probe a large area concurrently at different spectral wavelengths are perhaps the quintessential benefits of using airborne \spectralimaging~for remote sensing (\remotesensing). Food and agriculture \citep{ekanayake2018semi}, ecology \citep{zhang2021advances}, hydrology \citep{chen2020comparative}, and mineralogy \citep{grebby2014impact} are to name but a few of the prominent fields with applications of \remotesensing.

In mineralogy, \remotesensing~is widely used for `mineral-indication' \citep{yousefi2016mineral}: the process of identifying minerals. Remote sensing emphasizes the spectral and radiometric properties as opposed to the chemical and physical characteristics of minerals. Decades-long research in mineral-indication has created spectral libraries that can be used to identify and categorize minerals to produce coverage maps. This has paved the way to use \remotesensing~ techniques to emulate the process of lithological mapping \citep{black2016automated, feng2003topographic, grebby2011integrating} to improve the interpretation of the lithology of geological sites. In digital lithological mapping (\dlm), hyperspectral images (\spectralimage{H}s) are used due to large spectral feature availability, whereas multispectral images (\spectralimage{M}s) with narrowband sensors \citep{del2020evaluation} too are used. To be specific, \dlm~ has been performed using thermal infrared (TIR) data with broadly spaced multispectral sensors. Further, the Thermal Infrared Multispectral Scanner and the Advanced Space-borne Thermal Emission and Reflection Radiometer sensors have illustrated the usefulness of TIR data to differentiate a broad spectrum of minerals \citep{hubbard2005mineral, jiang2013lithological, ninomiya2005detecting}, especially silicates. The primary reason to use \spectralimage{M}s for \dlm~ is its inherent improved signal-to-noise ratio (SNR) from the broad spacing between spectral bands which mitigates the cross-channel interference in contrast to \spectralimage{H}s. Nonetheless, the use of \spectralimage{H}s is enabled with an algorithm that has superior noise performance owing to spectral richness as opposed to \spectralimage{M}s. Besides that, the geographical sites explored in these studies are solely constituted of minerals and are destitute of other environmental features such as trees, soil, sand, etc. which hinders the direct application of available laboratory-generated spectral signatures \citep{ismail2014rare,kruse2007regional,pour2019lithological} unlike for peninsulas.

To identify the mineral deposit around a peninsula, it is imperative to study the composition and identify the surface minerals in that area. Naturally, peninsulas are formed around mineral deposits with the intervention of tides and currents over time. As the peninsula forms, the nearby surface is populated by plants and mineral residues \citep{jia2020sedimentary, senaratne1982palaeogeographic}. When civilizations are built around these areas, though the original mineral deposits are likely to be preserved, new and alien substances will be introduced to the environment. Hence, the surface formation of a peninsula with a human population is different from a one without. Consequently, the spectral images of such sites will include signatures from elements other than minerals, so the use of unmixing techniques is hamstrung as the number of principal elements is imponderable. Alternatively, the existence of a collection of predetermined minerals can be detected with spectral libraries \citep{rajendran2018spectral} using criteria such as maximum likelihood classification \citep{cabral2018burned}, spectral angle mapping \citep{dennison2004comparison}, and spectral information divergence \citep{palsson2017neural} because the error of misclassification is distributed amongst several mineral types. But the same methods will be futile if to be used in the detection of a particular mineral because coalescing with other constituents could distort the spectral characteristics of the target mineral and cause it to diverge from the library signature. Furthermore, certain minerals that have a resemblance to the target mineral in terms of spectral properties might be confounded as the mineral of interest under single-target detection. Nonetheless, single-target mineral detection is more suitable for predictive modeling of the distribution of a mineral \citep{carranza2009predictive} in each location with no prior information related to the lithology. In addition, the same estimated distribution could be used as a precursor to sort site locations as to survey (i.e. places with limestone presence as returned by the method) and avoid (i.e. places with limestone absence as returned by the method) before an extensive geological survey.

The East coast of the Jaffna Peninsula of Sri Lanka is such a geographical site where the lithological data of the dominant mineral is unavailable. The Jaffna peninsula has been developed around a limestone deposit hence, it is considered as the dominant lithology. However, the limestone deposits are copious towards the West coast of Sri Lanka according to the findings of \cite{perera2020cement}. The study location selected in the work was in the vicinity of the limestone region but lacks information about the existence of limestone from previous studies. Further, it is improbable to excavate the entire survey site to determine the underground spread of the deposit, so the best practice is to collect samples from various pragmatically selected locations. The residues from the deposit will emerge to the surface as the peninsula progresses \citep{senaratne1982palaeogeographic} and the variation of the residues are useful to establish the spread. The composition of the soil has been studied from surveys, and survey points (i.e. samples collected locations) are used to construct contour plots for the variation of the mineral upon laboratory confirmations of minerals. Hence, the produced lithological map is a suboptimal representation of the mineral spread regardless of the tedious and meticulous sample collection process. Besides understanding the distribution of the dominant mineral around the region, detection of limestone is important for three major reasons,
(a) the development of the highly irregular soil bedrock contact,
(b) the frequent development of subsurface cavities with a possibility to behave as sinkholes, according to \cite{adams1984mapping}, and
(c) commercially crucial as an ingredient for the cement industry.

In this regard, an algorithm is proposed in this work for the single-target detection of limestone and to generate an abundance map that serves as a precursor to site selection for a geological survey with the Jaffna peninsula as the case study. The proposed algorithm incorporates \spectralimage{H}s owing to its superior discriminating power in terms of spectral features despite low SNR characteristics from cross-channel interference compared to \spectralimage{M}s. Furthermore, the proposed algorithm utilizes existing hyperspectral libraries for target mineral detection but incorporates stochastic cancellation principles to extract limestone signatures that inherit spectral properties of the pure mineral but are accounted for the spectral variation introduced by other constituents referred to as `residual impurities'. This is because, given the human interventions around this region, it is impossible to perform mineral-indication solely relying on spectral libraries since the mineral signature could be distorted \citep{alkharabsheh2013impact, uddin2013application} by foreign materials. Besides that, the surface mineral composition gradually changes in the area, therefore synthesized spectral signatures are inapplicable to classify the principal mineral: limestone. Therefore under these circumstances, the proposed algorithm's stochastic cancellation process which generates the site-specific signature is the only pragmatic viable alternative.

First, land cover mapping was performed as a pre-classification on the \spectralimage{H} to separate pixels that subsume spectral properties mostly of soil. Afterward, the correlation of the soil pixels with a library reference signature for limestone was calculated to create representative sets for site-specific signatures for limestone and residual impurities. Next, a site-specific limestone signature was extracted using stochastic cancellation to include second-order correlation properties of the library limestone signature. 
Before that, an estimated endmember signature for residual impruities was subracted from the signatures in the limestone representative set through a linear unmixing process to preclude undesired signatures containing correlated components with the library limestone signature.
Finally, the extracted endemic limestone signature was used to calculate the abundance values of limestone in soil pixels to generate a mineral map for limestone as the outcome of the algorithm. The proposed work facilitates the likelihood estimation of limestone existence in the survey area using \spectralimage{H} data. The benefit of this likelihood estimation is twofold:
(a) the mapping process is a precursor for the field survey and
(b) inference on the limestone availability.
Firstly, these two results could be used to select locations for sample collection, ergo will expedite the process and will mitigate specific difficulties such as determining an optimal number of survey points and locations, resource allocation, and observation of obscure sites. Secondly, the generated mineral map either could be used to juxtapose existing knowledge on lithology from surveys as a confirmation strategy or infer insights on sites that are assumed to be deprived of limestone.

The contributions of the proposed work can be enumerated as follows;
(a) extraction of a limestone signature with second-order correlation properties of the library reference signature but is adjusted for the spectral variation inherent to the survey site.
(b) soil pixel alignment based on the mineral purity level.
(c) generation of a mineral map with the site-specific limestone signature as a predictive model for the selection of sites to survey or avoid during the field survey.
(d) derivation of a smooth contour map for abundance extrapolation and estimation.
(e) mitigation of poor accuracy and overestimation from using generic laboratory signature for mineral presence detection.
(f) single-target mineral detection of limestone in the presence of different unknown minerals.

In section \ref{section: related work}, related work on the use of \spectralimage{H}s in mineralogy and endmember extraction techniques has been presented. Next, the extraction of the site-specific limestone signature and generation of the predictive mineral map preceded by pre-classification of pixels are delineated in section \ref{section: methodology}. Following, information on the field visit, collection of in-situ data, and laboratory validation tests are provided in section \ref{section: field study}. Afterward, the results of the algorithm and the field survey with a discussion about the results are presented in section \ref{section: results and discussion}.

\section{Related work}
\label{section: related work}
The development of hyperspectral sensors that are capable of producing images from hundreds of contiguous spectral channels improved the range and accuracy of remotely retrieved mineralogical information \citep{niranjan2016mapping, swamy2017remote} and surface composition \citep{zomer2009building}. These hyperspectral cameras either range from infrared to ultraviolet or are restrained to the TIR region. Recently, several airborne hyperspectral TIR instruments, including the Airborne Hyperspectral Scanner, ITRES Thermal Airborne Spectrographic Imagery, and the Spatially Enhanced Broadband Array Spectrograph System are in operation and have demonstrated the potential for mapping clays, sulfates, silicates, and carbonates \citep{vaughan2003sebass, zhang2018hyperspectral}. Recent studies in the lithologic mapping of peninsulas \citep{fiaschi2017complex, rowan2003lithologic} have been carried out using TIR data from airborne hyperspectral sensors with automated identification and quantification processes. Besides, owing to greater spatial resolution, \spectralimage{H}s available via satellites or airborne imaging systems, have been used to analyze minerals considering soil variability at different \remotesensing~ scales and to measure rock microstructures \citep{van2019measuring} as well. 
Further, the use of generic laboratory-generated spectral libraries has been the prominent method used in most digital lithological studies \citep{black2016automated, ekanayake2019mapping, feng2003topographic, grebby2011integrating, tziolas2020integrated} whereas generation of a site-specific signature has not been a focus. Moreover, the minerals of which the generic spectral signatures were used for lithological mapping were either assumed to be present in the site \citep{ninomiya2019thermal, yu2012towards} or known from an already conducted survey of the site \citep{pour2018mapping, xiong2011lithological}. Here, the former method could be error-prone as the signatures are not site-specific and the latter is only a validation procedure that neither enriches nor expedites the field survey \citep{rajan2019mapping}.

The extraction of endmember signatures is often addressed and is a discrete research area in the literature for \remotesensing~ unmixing. These studies make use of unsupervised unmixing techniques in \remotesensing~ to derive the endmember signatures and the abundance map for the lithological map. For example, orthogonal subspace projection methods \citep{cheng2015remote, li2015recursive,ren2000generalized} were used to classify spectral signatures following dimension reduction algorithms. Furthermore, accurate \spectralimage{H} classification has been performed via an improved version of the standard non-negative matrix factorization (NMF) algorithm incorporating fundamental notions of independence \citep{benachir2013hyperspectral, sun2017poisson}. Though existing techniques such as NMF-based unmixing \citep{rajabi2013hyperspectral,rathnayake2020graph,wang2016hypergraph} or autoencoder architecture based unmixing \citep{hua2020autoencoder,qi2020deep,ranasinghe2020convolutional} are superior at extracting the endmembers and estimating the corresponding abundances, these algorithms require the knowledge about the number of endmembers to extract which was not available in the first place. Besides, for an algorithm to find feasible survey locations for a particular mineral through single-target identification, such information is superfluous. Albeit, blind source separation algorithms such as Pixel-Purity-Index algorithm, Vertex Component Analysis, Independent Component Analysis are catered to extract source signals, lack of pure limestone pixels, and information on endmembers frustrate the use of these statistical and geometrical algorithms. Besides that, most of the study areas have had a mineral composition that allowed automated unmixing algorithms to extract endmembers accurately. 
Even though the endmember extraction with the above methods is viable for unpopulated geographical locations, in certain geographical regions human intervention \citep{barbosa2003otter, mendes2015multi, vila2016stakeholder} has created residual impurities that could make these methods unreliable.

According to the related works, on one hand, \dlm~ has been performed using spectral libraries of geographical regions that had a pure mineral composition, thereby the mineral signatures were more dominant than the spectral signature of other non-mineral constituents or impurities. Although, there is a limited number of works to improve the SNR of remote sensing images, detection of minerals under a strong influence of impurities is yet to be improved to the best of our knowledge.
On the other hand, estimation of the abundances for mineral mapping has been performed under multi-target detection with unmixing techniques. But, the mineral map generation for a target mineral has not been under scrutiny since these unmixing techniques are susceptible to noise in the image.
Hence, the proposed work addresses the gaps identified in the literature with the contributions presented in \ref{section: introduction} for site-specific endmember generation to estimate the abundances of a particular mineral under single-target detection.

\section{Methodology}
\label{section: methodology}

\subsection{Pre-classification}
\label{subsection: pre-classification}
To reduce misclassification of other minerals as `limestone' and for computational efficiency, the HSI was pre-classified into subcomponents. For this, sub-component analysis \citep{ekanayake2019mapping} was used in combination with vertex component analysis (\vca) \citep{nascimento2005vertex}. As suggested in \cite{ranasinghe2019hyperspectral}, pixels were categorized into four classes: soil, water, vegetation, and sand. First, candidate endmembers were extracted using \vca~ for the four classes. Then, each HSI pixel was normalized by the $L\textsubscript{2}$-norm to construct the respective directional vector because the direction of each vector characterizes the spectral behavior of the pixel. Similarly, the candidate endmembers too were normalized for consistency. Subsequently, the Euclidean distance to each normalized candidate endmember was calculated for each pixel. The goal was to obtain a similarity metric with the four candidate endmembers for each pixel. Thus, the reciprocal of the distance metric was used for the similarity measure which was defined and computed as,

\begin{align}
    \label{equation: similarity equation}
    \gamma_\text{m}^\text{i} &= \frac{\sfrac{1}{\Vert\mathbf{u_m}-\mathbf{r_i}\Vert_2}}{\sum_c \sfrac{1}{\Vert\mathbf{u_m}-\mathbf{r_c}\Vert_2}}
\end{align}
where, $\mathbf{r_i}\text{ and } \mathbf{u_m}$ denote the spectral signatures of the i\textsuperscript{th} reference vector and the m\textsuperscript{th} pixel respectively, hence the notation of $\gamma_\text{m}^\text{i}$ for the similarity between m\textsuperscript{th} and i\textsuperscript{th} reference signatures. The summation therein is taken over all the candidate endmembers.
The pixels were classified using a threshold for the largest percentage affinity ($\gamma_\text{m}^\text{i}$) recorded for each pixel according to \ref{equation: similarity equation}. That, any pixel with a $\gamma_\text{m}^\text{i}$ higher than $0.5$ was labeled as a pixel of the respective i\textsuperscript{th} class while the pixels with all $\gamma_\text{m}^\text{i}$s less than 0.5 were disregarded.

\subsection{Alignment of soil pixels for purity}
\label{subsection: alignment of soil pixels for purity}
Mineral identification was performed on the pixels classified under the soil class. However, the algorithm proposed in \cite{ekanayake2019mapping} requires \textit{a priori} knowledge regarding the composition of the soil to construct a representative spectral signature and the laboratory references for every major mineral compound that is assumed to exist. Nonetheless, it is not necessary to have information about the mineral composition of the soil as such reference collection is immaterial for an algorithm with an application of identifying probable survey locations for an earmarked mineral. Hence, a laboratory reference for limestone suffices for the initial identification of the availability, and the required spectral signature is available on USGS spectral library. But, the number of spectral bands of the limestone reference signature was greater than that of the satellite's detector, therefore these additional spectral bands along with low SNR bands of the detector were removed from the original reference signature. Consequently, the modified reference signature had a spectral band count equal to that of the HSIs prior to dimension reduction. Next, correlation factor analysis (CFA) was performed on soil pixels using equation

\begin{equation}
    \label{equation: pearson correlation calculation}
    r_k = \frac{\sum_{i=1}^b(x_{k,i}-\frac{1}{b}\sum_{j=1}^b x_{k,j})(s_i-\frac{1}{b}\sum_{j=1}^b s_j)}{\sqrt{{\sum_{i=1}^b}(x_{k,i}-\frac{1}{b}\sum_{j=1}^b x_{k,j})^2(s_i-\frac{1}{b}\sum_{j=1}^b s_j)^2}}
\end{equation}
where, \(r_k\), \(x_{k,n}\), \(s_n\), and \(b\) represent the Pearson correlation coefficient between \(k^{th}\) pixel and the reference signature, value of the \(n^{th}\) spectral band of the \(k^{th}\) pixel, value of the \(n^{th}\) spectral band of the laboratory limestone signature, and the number of spectral bands, respectively.

\begin{figure*}[ht!]
    \centering
    \begin{subfigure}[t]{0.45\textwidth}
        \centering
        \includegraphics[width=1.0\textwidth]{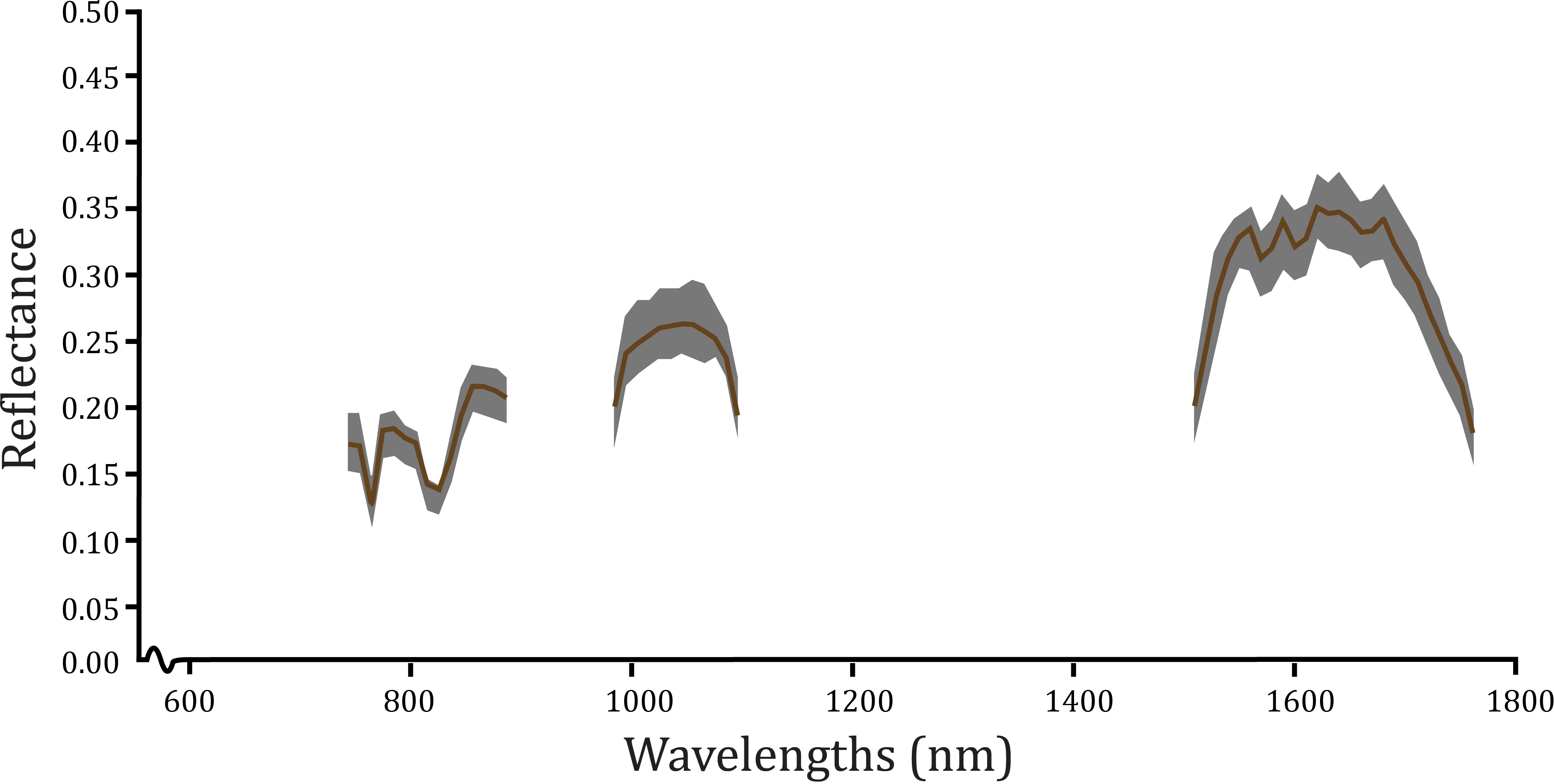}
        \caption{}
        \label{figure: limestone subclass signature with variability}
    \end{subfigure}
    ~ 
    \begin{subfigure}[t]{0.45\textwidth}
        \centering
        \includegraphics[width=1.0\textwidth]{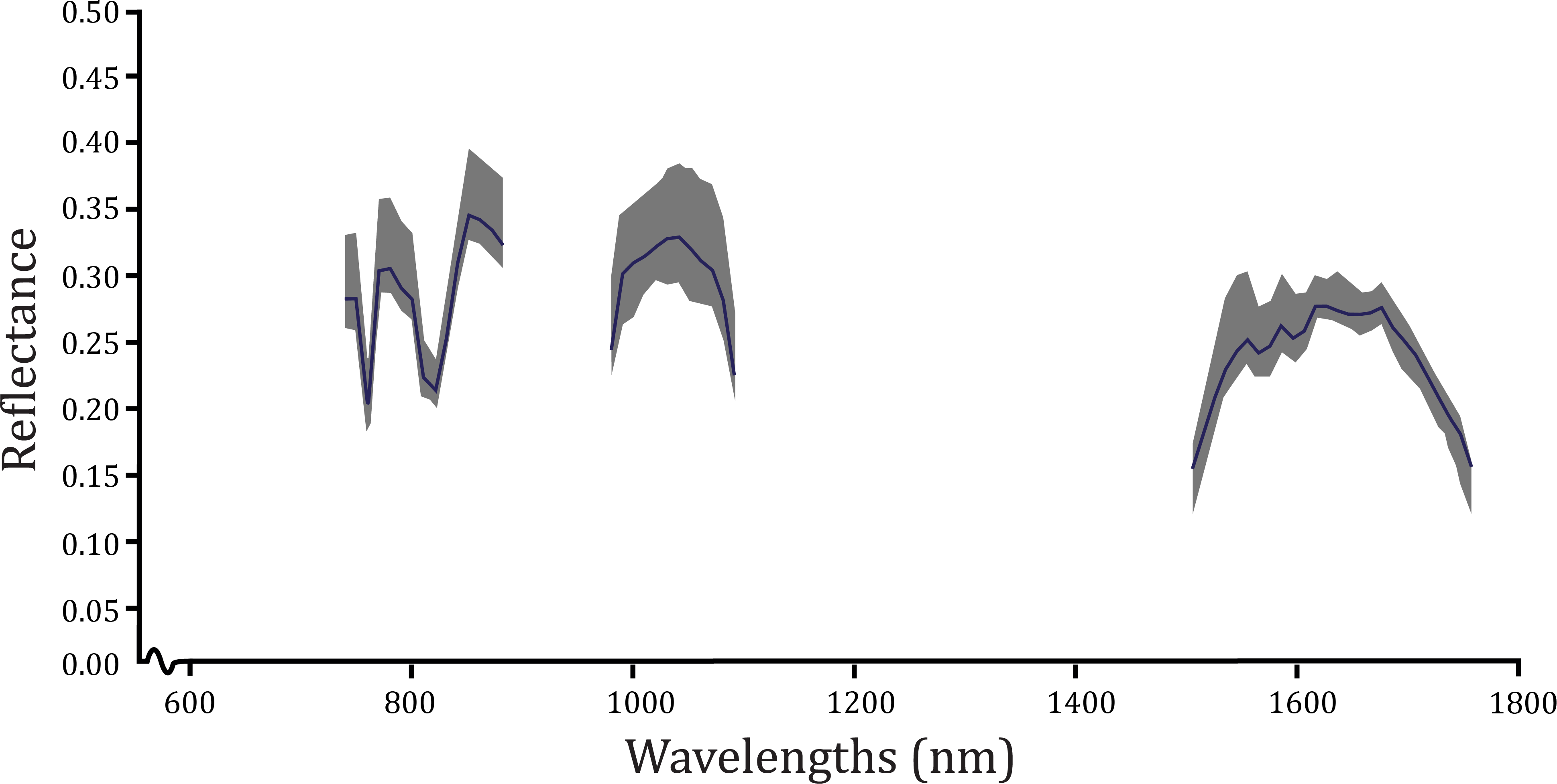}
        \caption{}
        \label{figure: residue subclass signature with variability}
    \end{subfigure}
    \caption{Representative endmember and variability in class signatures for subclass: (a) limestone (b) residual impurities}
    \label{figure: endmembers for subclasses with variability}
\end{figure*}

After the coefficients were computed for each of the soil class pixels, two sub-classes, namely, limestone and residual impurities were established for the calculation of relative availability (\relativemeasure) of limestone with a representative set of signatures for each sub-class. For the selection of representative signatures, first, an endmember for each sub-class was found from the soil class pixels using the \vca~ method. In Fig. \ref{figure: endmembers for subclasses with variability} the endmember of each class is given along with the variability in the representative set. However, neither the extracted limestone endmember is the most correlated, nor the residual impurity endmember is the least correlated representation with the library reference. But the correlation of each endmember provides an upper and a lower bound for the most correlated and least correlated pixels, respectively. Ergo, the correlation of each endmember was calculated according to \ref{equation: pearson correlation calculation} following CFA. Next, the pixels that had a correlation coefficient higher than that of the endmember for the limestone sub-class were considered as the representative set for that sub-class, and those with a correlation lower than that of the endmember for the residual impurity class were considered as the representative set for that class. Following, the endmember of each class was set as the mean of their representative set as opposed to the extracted endmember from \vca~ to incorporate the correlation characteristics with the reference limestone signature.

Before, extracting the endmembers it is important to investigate the variation of the correlation between soil pixels and the limestone reference signature as traversed from the least correlated set to the most correlated set along the soil signature manifold. For this, we consider the \relativemeasure~of limestone, a normalized metric to measure the amount of limestone in a given soil pixel compared to the mean of the most correlated set. In calculating the RA values, the manifold of the soil pixels was oriented using FDA to have a monotonous variation in the correlation with the limestone reference as traversed between the sub-classes. The use of FDA with two classes produces an eigen-direction along which the correlation is monotonically changing since the least-correlated and most-correlated sub-classes are on the opposite extremes along this eigen-direction. The FDA algorithm was applied to the dataset according to \ref{equation: fisher discriminant} and the transformation matrix was constructed from the resulting eigenvectors as below:

\begin{equation}
\begin{split}
    \label{equation: fisher discriminant}
    \boldsymbol{\mu_c} &= \frac{1}{N_c}\sum_{m=1}^{N_c}\mathbf{x_m} \\
    \mathbf{S_b} &= (\boldsymbol{\mu_{l}} - \boldsymbol{\mu_{r}})(\boldsymbol{\mu_{l}} - \boldsymbol{\mu_{r}})^\top \\
    \mathbf{S_w} &= \sum_{m=1}^{N_l}(\mathbf{x_{l_m}} - \boldsymbol{\mu_{l}})(\mathbf{x_{l_m}} - \boldsymbol{\mu_{l}})^\top \\
    &\hspace{5mm} + \sum_{m=1}^{N_r}(\mathbf{x_{r_m}} - \boldsymbol{\mu_{r}})(\mathbf{x_{r_m}} - \boldsymbol{\mu_{r}})^\top \\
    \mathbf{S_b\,v} &= \lambda\,\mathbf{S_w\,v}\\
\end{split}
\end{equation}
In \ref{equation: fisher discriminant}, $\mathbf{S_b} \text{, }\mathbf{S_w} \text{, }\lambda \text{, and }\mathbf{v}$ represent the between class correlation matrix, within-class correlation matrix, Eigenvalue, and corresponding Eigenvector. Further, $\boldsymbol{\mu_c} \text{, }{N_c} \text{, and }\mathbf{x_{c_m}}$ represent the mean spectral signature, number of pixels, and vector representation of soil pixels from each class. 

\begin{figure*}[h!]
    \centering
    \includegraphics[width=0.5\textwidth]{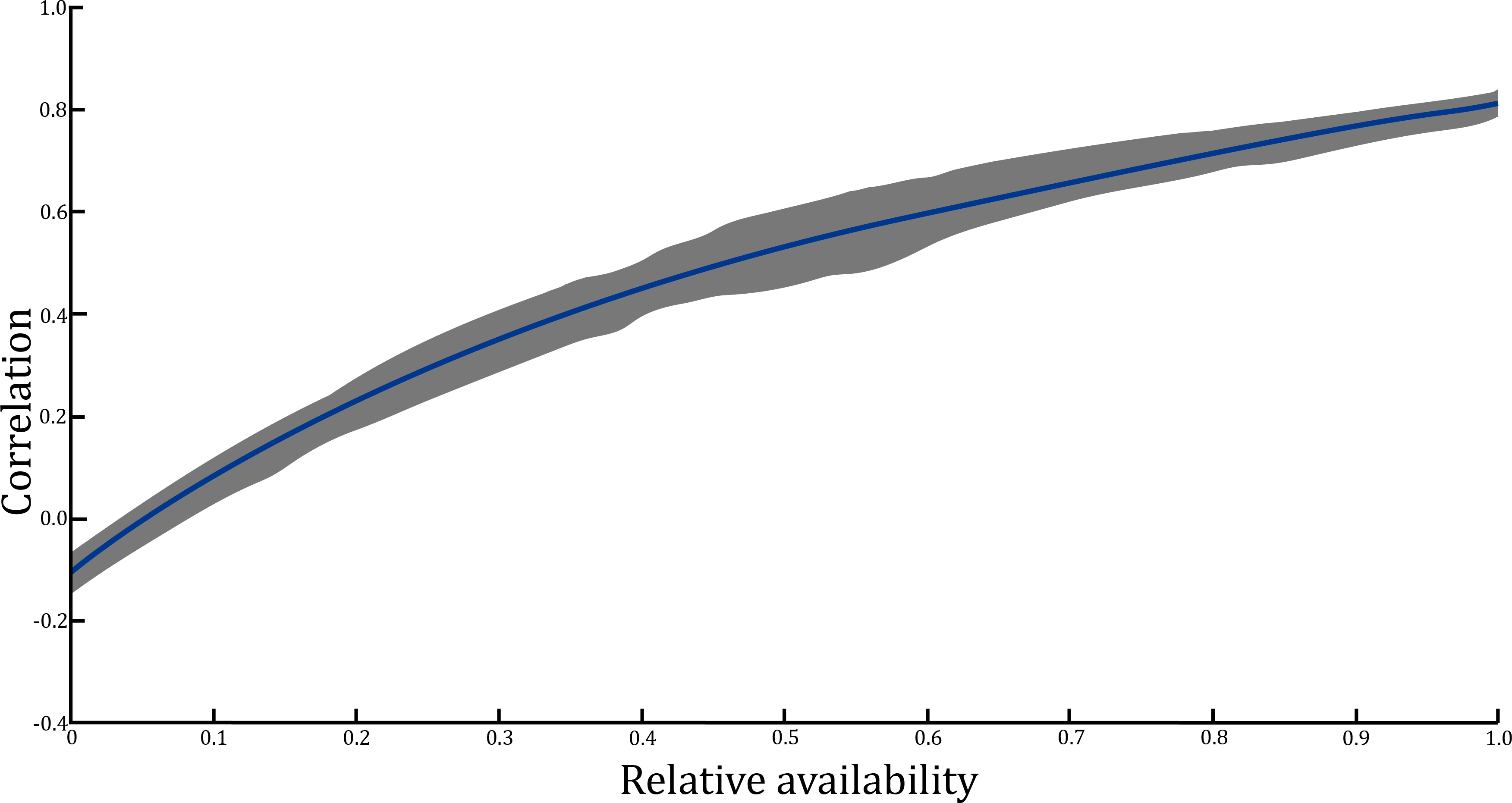}
    \caption{Variation of correlation with limestone library signature with relative availability}
    \label{figure: correlation with relative availability}
\end{figure*}

The eigen-direction was found through an eigen-decomposition of the matrix $\inv{\mathbf{S_w}}\text{\,}\mathbf{S_b}$. Following, the mean spectral signatures of the two sub-classes were projected onto the eigen-direction along which the correlation variation is monotonous. Similarly, the pixel vectors of the soil class were also transformed to this new subspace and the \relativemeasure~of limestone in pixels was calculated in comparison to the transformed reference signatures. Then, the distance to both the mean signatures from the pixel was calculated for each pixel along the eigen-direction. The reciprocal of each distance was taken as a measurement for the affinity and the \relativemeasure~of limestone was defined as,
\begin{equation}
    \label{equation: relative availability of limestone}
    \begin{split}
    \mathbf{\textit{\parbox{18mm}{Relative availability}}} &= \frac{\textit{\parbox{55mm}{similarity with the limestone reference}}}{\textit{\parbox{55mm}{total of similarities with references}}}\\
    &= \frac{\sfrac{1}{d_l}}{\sfrac{1}{d_l}+\sfrac{1}{d_{ri}}}\\
    &= \frac{d_{ri}}{d_{l}+d_{ri}}
    \end{split}
\end{equation}
where $d_{l}$, and $d_{ri}$ are the distances from the pixel to the mean signatures of the limestone and residual impurities sub-classes, respectively. Since the formation given in the second step of \ref{equation: relative availability of limestone} could lead to numerical instabilities for small $d_{l}$ values, the form given in the third step was used in the implementation. The RA of limestone is a measure for the amount of limestone contained in a given pixel compared to the pixels with a higher probability to subsume limestone residues, and the values for \relativemeasure~range from 0-1 as illustrated in Fig. \ref{figure: correlation with relative availability}. According to Fig. \ref{figure: correlation with relative availability}, the variance in the correlation is minimal towards the extremes of the manifold, which permits the use of mean signatures for the limestone endmember extraction. Though the absolute value of limestone is not given by \ref{equation: relative availability of limestone}, the calculated value provides an approximation for the lower availability values and is useful in the endmember extraction of the algorithm in section \ref{subsection: extraction of site-specific limestone endmember}.

\begin{figure*}[h!]
    \centering
    \includegraphics[width=0.5\textwidth]{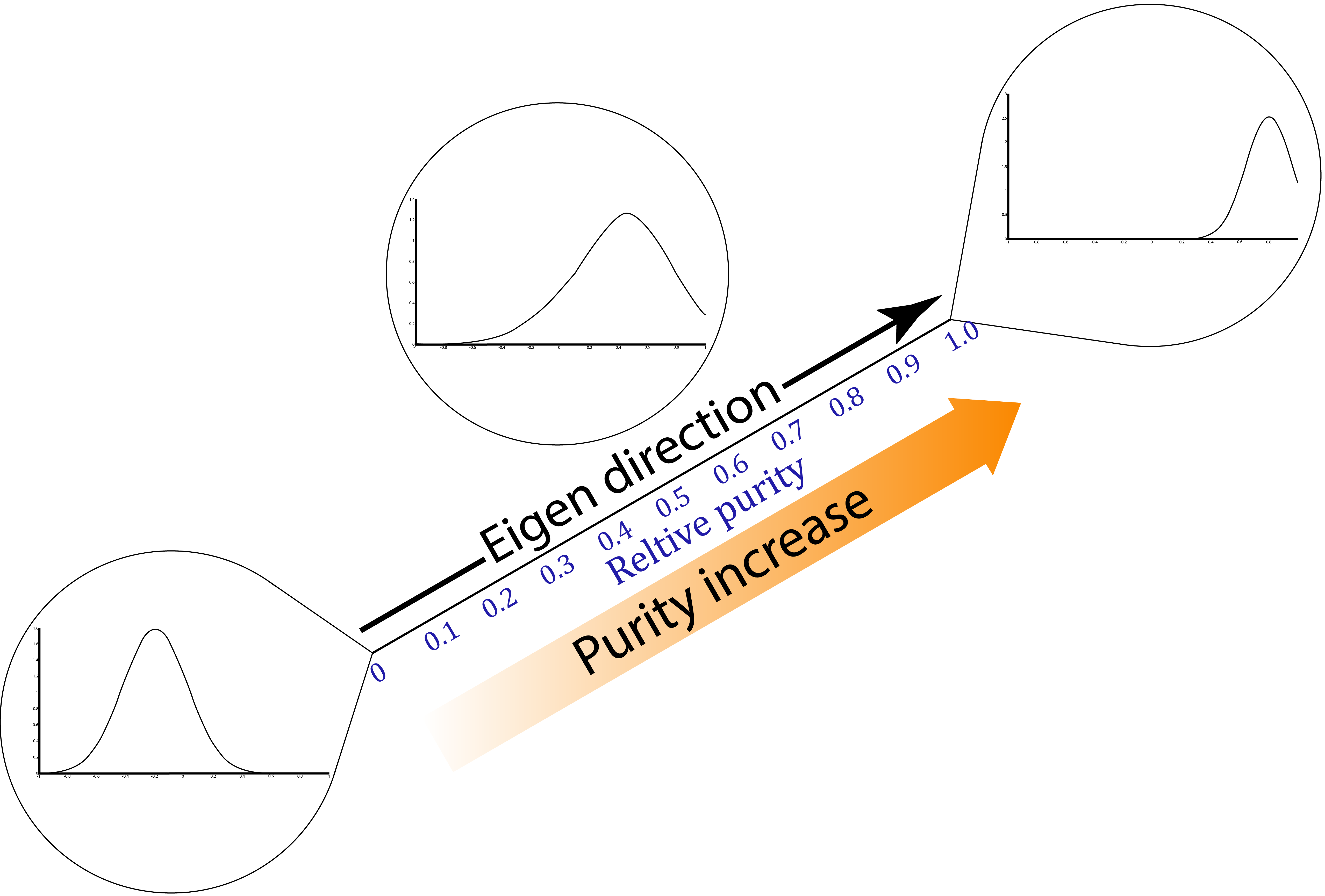}
    \caption{Alignment of soil pixels along the eigen-direction for relative limestone purity}
    \label{figure: pixel purity alignment}
\end{figure*}
Though the alignment of pixels is an integral part of the site-specific mineral extraction, the intermediate result has provided a method to find an optimal direction to align a given set of soil pixels along which the purity of the target mineral increases. As illustrated in Fig. \ref{figure: pixel purity alignment}, the eigen-direction extends from the residual impurity subclass to the limestone (i.e. target mineral) subclass. To expound on the purity scale, the distribution of correlation with the laboratory signature is given for pixels with similar purity values at three locations on the scale and the shift in the distribution from left to right corroborates the mineral purity variation.

\subsection{Extraction of a site-specific limestone endmember}
\label{subsection: extraction of site-specific limestone endmember}
An adaptive algorithm based on the Least-Mean-Square algorithm was proposed to extract the endmember for limestone from the hyperspectral dataset. Further, a Wiener filter arrangement for inverse modeling allows using the available limestone signature for endmember extraction. In theory, when a signature is fed to the Wiener filter, the correlating portion of that signature with the desired signature (laboratory signature) will be the filter output. Thereby, the site-specific limestone signature which constitutes the correlated portion of soil pixel with the limestone reference, is still recoverable in the presence of other constituents, assuming the second-order statistics of the pixels with higher RA are not severely distorted by other constituents. Since pixels with higher RA are more correlated with the desired signature as observed in Fig. \ref{figure: correlation with relative availability}, the use of the Wiener filter is conducive to extract an endmember with a better correlation with the limestone laboratory signature that is endemic to the survey site.

Accordingly, for the endmember extraction process, the pixels with a higher \relativemeasure~values were used as their high SNRs are desirable for the endmember extraction process. But these pixels contain interference from residual impurities. With the Wiener filter arrangement for inverse modeling, the correlated section of the residual impurity signature with the laboratory limestone signature (desired signal) will appear at the output of the filter. To remove the effects of residual impurities at the output of the Wiener filter, the effect of the residual impurities had to be removed from the pixel signatures with high \relativemeasure~values. Though the spectral signature of residual impurities is volatile in the spatial domain, an approximation is useful for the endmember extraction and its effects are negligible on the pixels with a high \relativemeasure~as well. Of course, the volatility of the impurity signature could be lowered with site-specific impurity signatures, which is a recursion of the proposed algorithm on a smaller survey region. To construct a signature for the residual impurities, pixels with the smallest \relativemeasure~were used as their correlation with laboratory reference is also low.
Assuming a linear mixture model, the soil spectral signature ($\mathbf{s}$) can be represented as,
\begin{equation}
    \begin{split}
    \label{equation: formation of signature}
        \mathbf{s} &= \{\alpha\mathbf{l} + \beta\mathbf{r}~|~ 0\leq\alpha,\,\beta \leq 1;~ \alpha + \beta = 1\}\\
    \end{split}
\end{equation}
using the signatures of the two endmembers. In \ref{equation: formation of signature} and \ref{equation: residue amount}, the spectral signature and its respective abundance are denoted by $\mathbf{l} \text{ and } \alpha$ for the limestone signature and $\mathbf{r} \text{ and } \beta$ for the residual impurity signature. Inner product operation and $L\textsubscript{2}$-norm are represented by $\langle~\cdot~\rangle$, and $\Vert~\cdot~\Vert$. Then, with the abundance sum-to-one constraint imposed in \ref{equation: formation of signature}, the equation can be rearranged as follows,
\begin{equation*}
    \begin{split}
        \mathbf{s} &= (1-\beta)\mathbf{l} + \beta\mathbf{r} \\
        (\mathbf{s - l}) &= \beta(\mathbf{r - l})
    \end{split}
\end{equation*}
to estimate the effect of residual impurities using,
\begin{equation}
    \label{equation: residue amount}
        \beta = \frac{\langle(\mathbf{r - l}),(\mathbf{s - l})\rangle}{{\Vert\mathbf{r - l}\Vert}^2}
\end{equation}

The library spectral signature for limestone was used to evaluate $\beta$ values and the spectral signature of the pixels with high \relativemeasure~were modified by deducting the impurity portion. The modified signal was fed as the input to the Wiener filter arrangement and theoretically, a first-order filter arrangement should suffice to model the inverse or the unmixing process. However, consecutive spectral bands could be correlated in signatures, therefore a higher filter order was chosen for the filter. But, during preprocessing certain spectral bands were removed and the continuity in the wavelengths was interrupted. As a result, the spectral response was separated into discrete regions: 742--885\,nm, 983--1094\,nm, and 1508--1760\,nm. Considering the filter order the cross-correlation between these regions was considered to be negligible, so the signature extraction was performed separately in these spectral regions.

The cancellation of the impurity signature from the high RA pixels and the extraction of the site-specific limestone endmember was performed using the following initial steps:
\begin{enumerate}
    \item The ensemble average of the soil pixels from the sub-class with low RA values was computed and selected as the impurity representative indigenous to the survey site.
    \item The rescaled impurity representative was removed from each pixel in the limestone subclass according to the corresponding $\beta$ value of each pixel. 
    \item Then the ensemble average of the modified limestone sub-class signatures was set as the representative of that sub-class.
    \item The average value of the limestone sub-class representative was removed from it to construct a zero-mean input signature to the Wiener filter.
    \item Step 4 was repeated with the library spectral signature for limestone to create a zero-mean desired signature.
    \item Signatures from step 4 and step 5 were used as inputs to the filter arrangement to obtain the site-specific limestone signatures as the output of the filter.
\end{enumerate}

\begin{figure*}[h]
    \centering
    \includegraphics[width=0.8\textwidth]{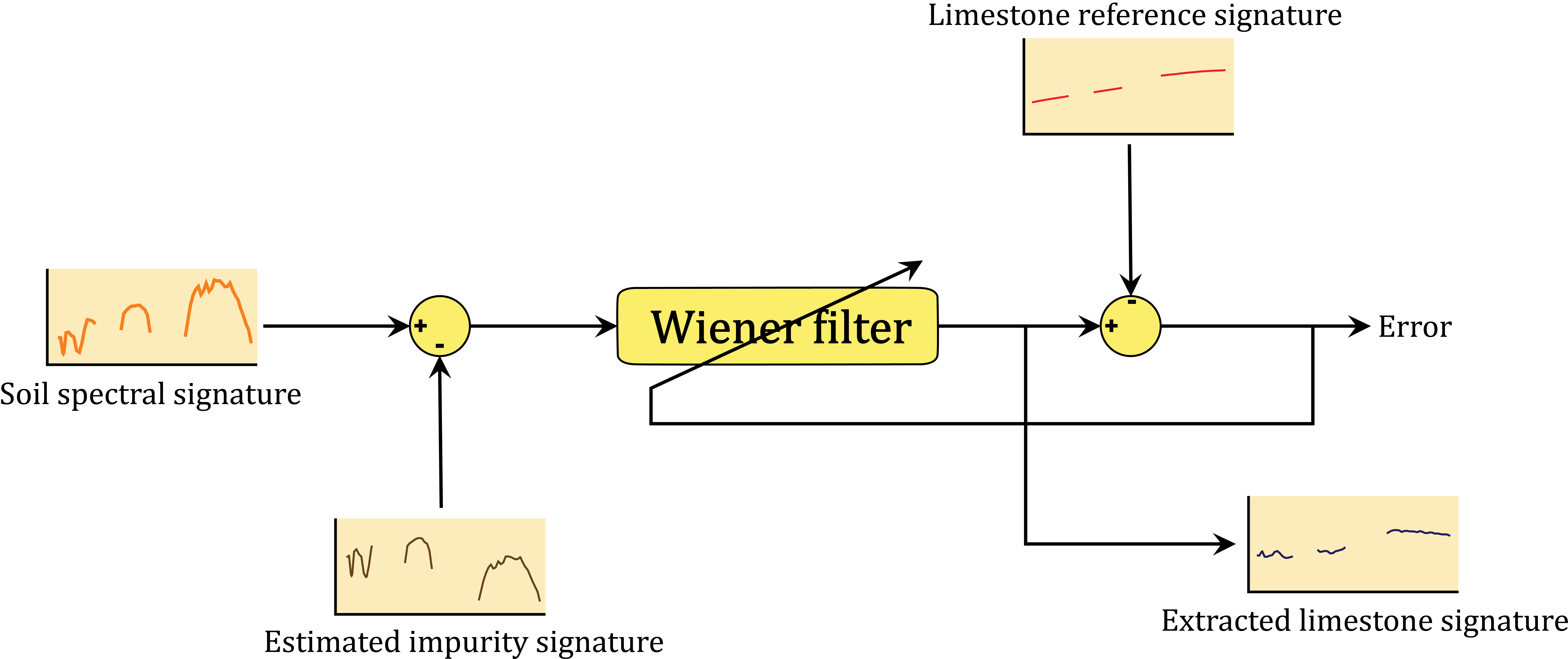}
    \caption{Wiener filter arrangement for the endmember extraction of limestone}
    \label{figure: wiener filter arrangement}
\end{figure*}
\begin{figure*}[h]
    \centering
    \includegraphics[width=\textwidth]{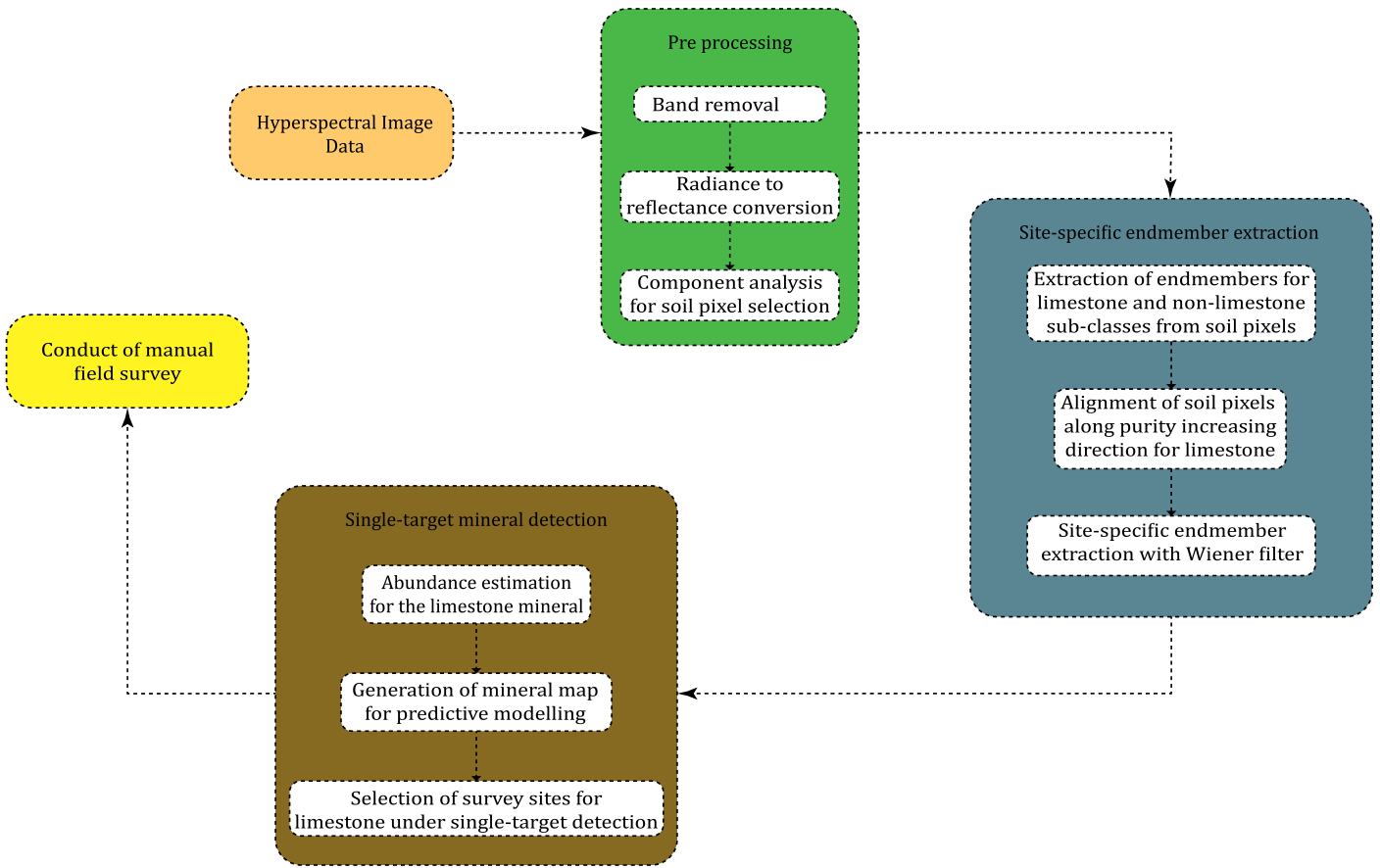}
    \caption{Flow chart of the proposed method}
    \label{figure: flow chart of the proposed method}
\end{figure*}

Since the objective of the optimization is to minimize the error between the output and the desired signal, the output of the Wiener filter will be the extracted limestone signature specific to the site. The site-specific endmember extraction process with the Wiener filter arrangement is provided in Fig. \ref{figure: wiener filter arrangement}.
Afterward, the soil spectral signatures were refined using the trained Wiener filter to extract the corresponding limestone signature component.
This extraction process further improves the separation between signatures of the least and most relatively available classes. Since spectral signatures are adjusted according to their composition, the two newly generated references were considered as more accurate references for limestone class and residual impurities class. Next, the abundance value for limestone was calculated using \ref{equation: limestone abundance estimator}. Similar to the derivation of \ref{equation: residue amount}, assuming a linear mixing model and the abundance sum-to-one constraint, the abundance ($\alpha$) of limestone can be calculated as,
\begin{equation}
    \label{equation: limestone abundance estimator}
    \alpha = \frac{\langle(\mathbf{r - l),(\mathbf{s - r})\rangle}}{\Vert\mathbf{r - l}\Vert_2^2}
\end{equation}
with the usual notation as introduced in \ref{equation: formation of signature}. The flow chart of the proposed algorithm is given in Fig. \ref{figure: flow chart of the proposed method}.

\section{Field study and remote sensing data}
\label{section: field study}

\subsection{Geological background and remote sensing data of the study location}
\label{subsection: peninsula history and remote sensing data}
\begin{figure*}[ht!]
    \centering
    \begin{subfigure}[t]{0.9\textwidth}
        \centering
        \includegraphics[width=1.0\textwidth]{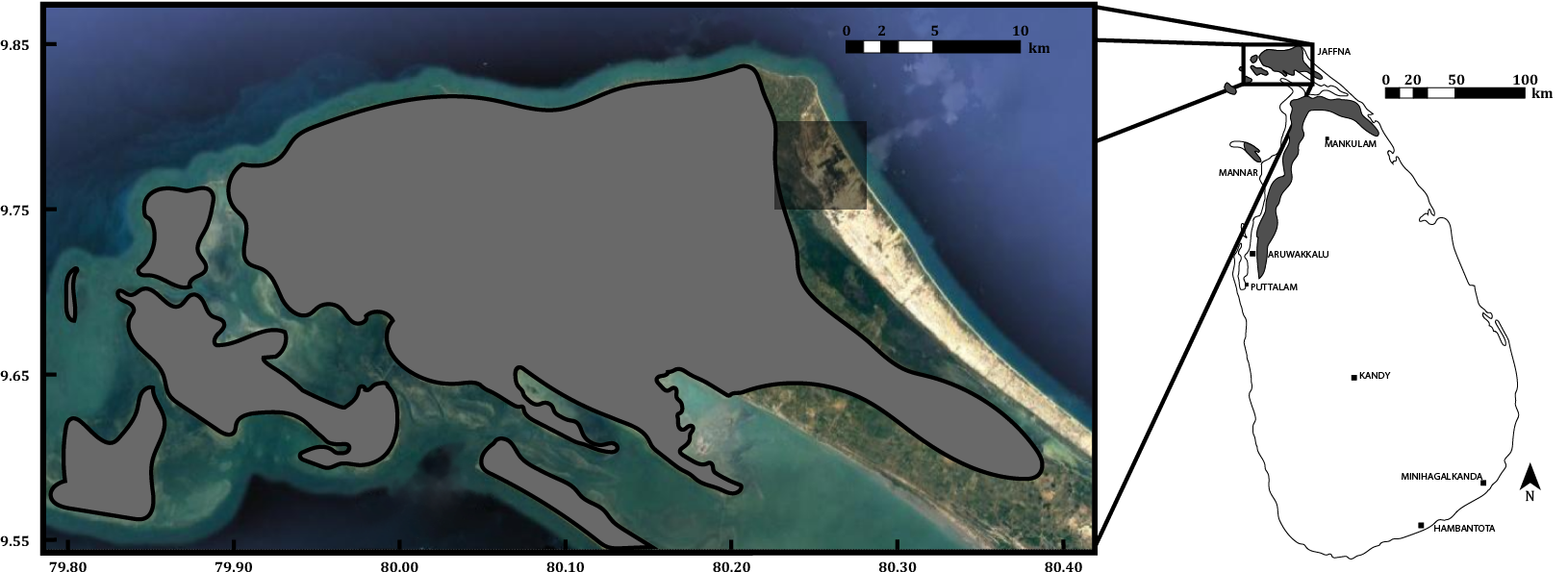}
        \caption{}
        \label{figure: study location with deposit distribution}
    \end{subfigure}
    \medskip 
    \begin{subfigure}[t]{0.9\textwidth}
        \centering
        \includegraphics[width=1.0\textwidth]{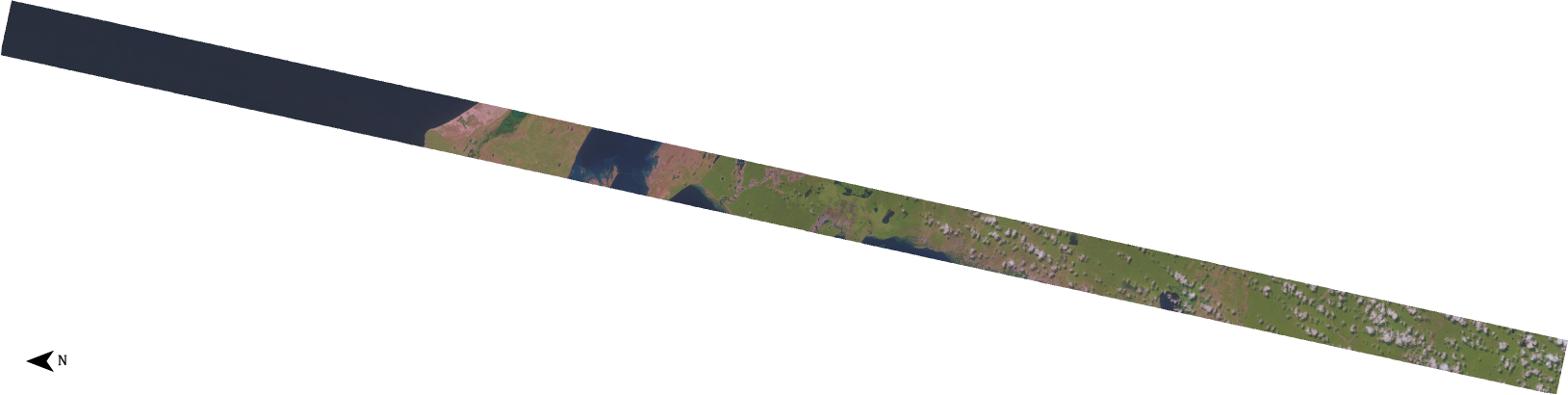}
        \caption{}
        \label{figure: hsi image strip}
    \end{subfigure}
    \caption{Satellite image of the (a) geographical location of the study area (black) and (b) remotely sensed image stripe}
    \label{figure: remote sensing images}
\end{figure*}

The island of Sri Lanka is a southern continuation of the shield area of Peninsular India. It is assumed that the island rotated a few degrees with respect to India without hindering the sedimentation process in the Cauvery Basin and adjacent shelf region of northwest Sri Lanka \citep{katz1978sri}. Stable shallow water conditions that continued until recent times have contributed to the deposit of Oligocene, Miocene, Pliocene, and Pleistocene continental shelf sandstones and limestones. The thick Jaffna Limestone is the dominant rock type that underlies the whole of the Jaffna Peninsula \citep{senaratne1982palaeogeographic} and the surrounding islands. Formerly, Jaffna was an island composed of Miocene limestone and the island was joined to the mainland by a spit formed of sediments brought by currents. These sediments were then carried to the eastern and northern coasts of the mainland and subsequently to the lagoon. Further, over time, the spits produced complex forms and some joined to form compound spits. The study location of this article given in Fig. \ref{figure: study location with deposit distribution} is a paleo-spit formed in the northeastern side of the peninsula which constitutes the shoreline. In Table \ref{table: study area cordinates} the corner locations of the study area are given.

\begin{table}[h]
    \centering
    \captionsetup{labelsep=newline, width=\columnwidth}
    \caption{Co-ordinates of the study area for the algorithm}
    \begin{tabularx}{\columnwidth}{ 
   >{\raggedright\arraybackslash}X 
   >{\centering\arraybackslash}X
   >{\centering\arraybackslash}X}
        \hline\hline
        Corner description & Latitude & Longitude \\
        \hline
        Upper left & 9.798306 & 80.229095 \\
        Upper right & 9.798306 & 80.281491 \\
        Lower left & 9.745226 & 80.229095 \\
        Lower right & 9.745226 & 80.281491 \\
        \hline
    \end{tabularx}
    \label{table: study area cordinates}
\end{table}

The \spectralimage{H} used in the study was captured by the Hyperion sensor attached to NASA's Earth Observing-1 satellite. The true-color image stripe of the \remotesensing~ data used for the analysis is given in Fig. \ref{figure: hsi image strip}. To reduce the computational time of the analysis, the algorithm was applied on a selected region as shown in Fig. \ref{figure: study location with deposit distribution} and described in Table \ref{table: study area cordinates}. Radiance values of the pixels were converted using \ref{equation: reflectance conversion equation} as below,

\begin{equation}
    \label{equation: reflectance conversion equation}
    \rho_\lambda = \frac{\pi\times L_\lambda\times d^2}{ESUN_\lambda\times\cos{(\theta_s})}
\end{equation}
with the notation 
\begin{tabbing}
$ESUN_\lambda$\=\kill
$\rho_\lambda$ \>: pixel reflectance at the wavelength $\lambda$\\
$L_\lambda$ \>: pixel radiance at the wavelength $\lambda$\\
$d$ \>: earth-sun distance in astronomical units\\
$ESUN_\lambda$ \>: exo-atmospheric solar irradiance\\
$\theta_s$ \>: solar zenith angle
\end{tabbing}
proposed in \cite{kokaly2017usgs}. Ancillary data such as mean solar exo-atmospheric irradiance for each band, and earth-sun distance in astronomical units for a set of days of the year are available in the USGS database. The geometrical parameters used for the conversion are given in Table \ref{table: conversion parameters}. 

\begin{figure*}[ht!]
    \centering
    \begin{subfigure}[t]{0.45\textwidth}
        \centering
        \includegraphics[width=1.0\textwidth]{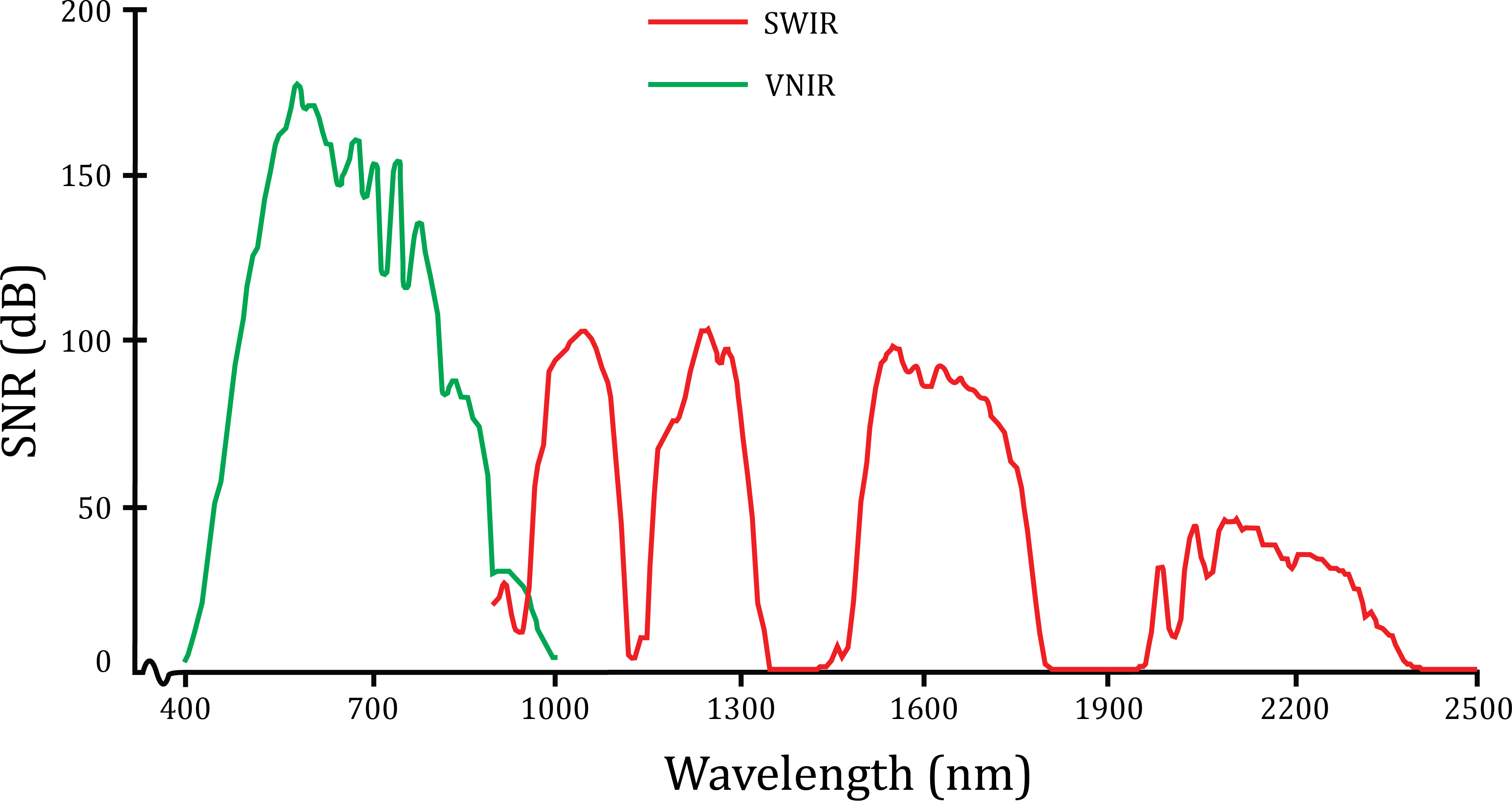}
        \caption{}
        \label{figure: snr variation of the sensor}
    \end{subfigure}
    ~ 
    \begin{subfigure}[t]{0.45\textwidth}
        \centering
        \includegraphics[width=1.0\textwidth]{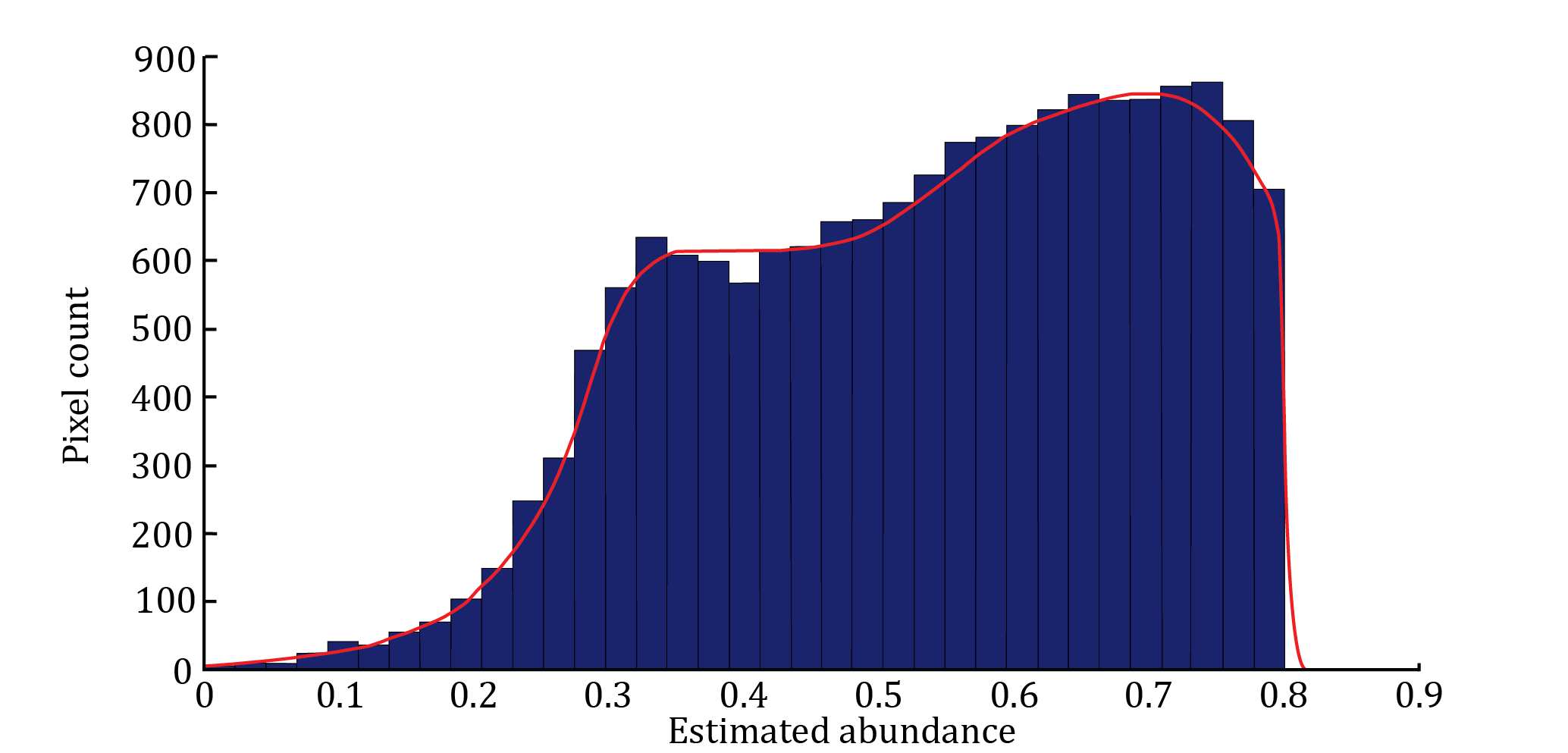}
        \caption{}
        \label{figure: abundance distribution of limestone}
    \end{subfigure}
    \caption{(a) Signal-to-noise ratio variation of the Hyperion sensor, and (b) distribution of the estimated limestone abundances for the site}
    \label{figure: snr and abundance distribution}
\end{figure*}

On the other hand, the laboratory spectral signature used in the algorithm has a higher SNR across the spectral range compared to the \spectralimage{H}s. Since the monochrome images from the low SNR bands would not contain useful information and could ultimately frustrate the performance, the images \citep{swayze2003effects} of the spectral bands with an SNR less than 50\,dB were removed from both the hyperspectral dataset as well as laboratory reference signature. The variation of SNR for the spectral bands are illustrated in Fig. \ref{figure: snr variation of the sensor}.

\begin{table}[h]
    \centering
    \captionsetup{labelsep=newline, width=\columnwidth}
    \caption{Parameters for the radiometric conversion of the satellite images}
    \begin{tabularx}{\columnwidth}{ 
   >{\raggedright\arraybackslash}X 
   >{\centering\arraybackslash}X}
        \hline\hline
        Parameter & Value \\
        \hline
        Earth-sun distance & 1.012 au \\
        Solar zenith angle & 28.5461 \\
        \hline
    \end{tabularx}
    \label{table: conversion parameters}
\end{table}

\subsection{Location selection and sample collection}
\label{section: sample collection}

\begin{figure*}[h!]
    \centering
    \includegraphics[width=0.9\textwidth]{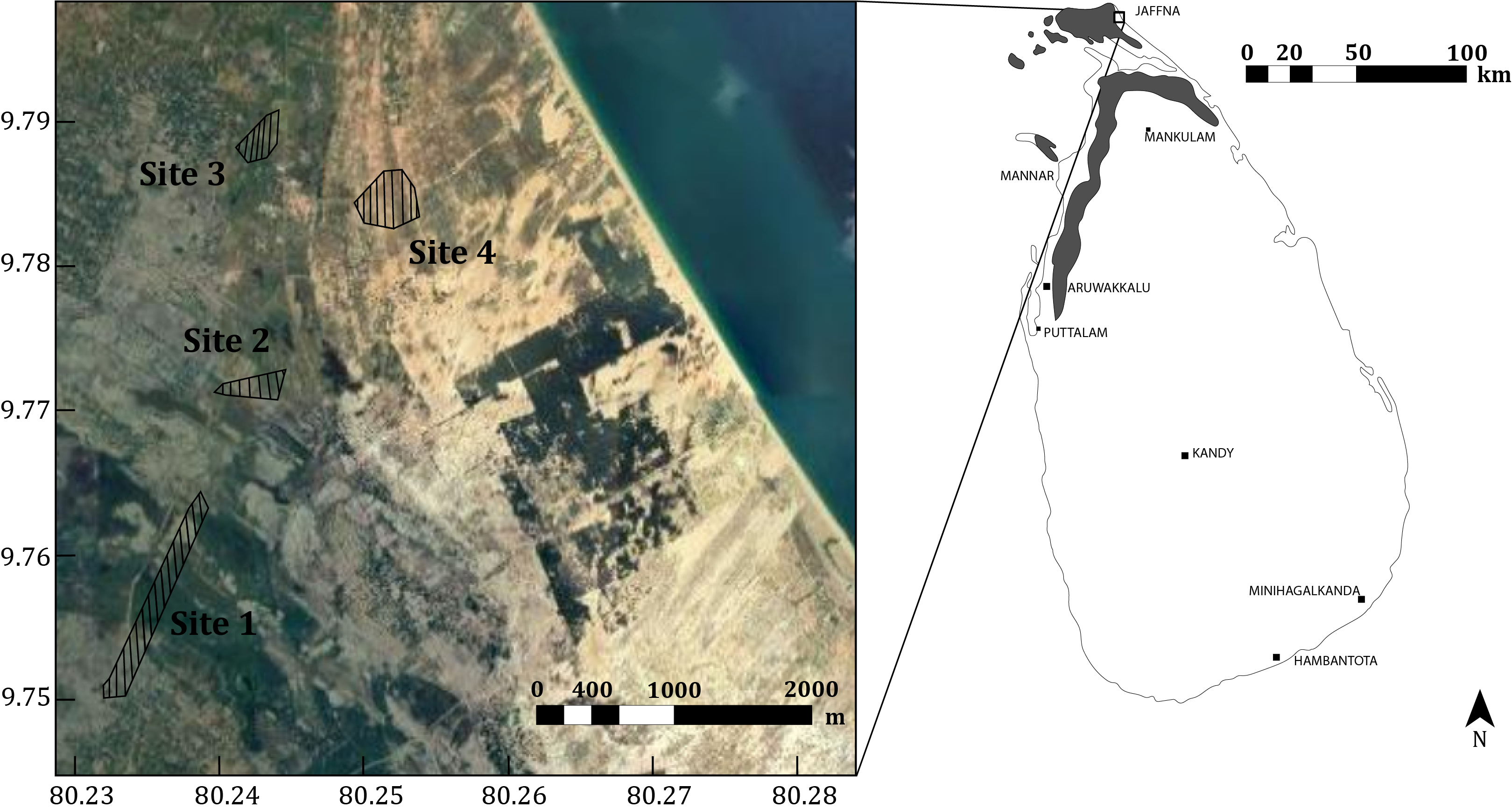}
    \caption{Selected site locations (in black) for the field survey and in-situ data collection}
    \label{figure: site selection}
\end{figure*}

Following the preprocessing of the \remotesensing~ data, the proposed algorithm was tested on the hyperspectral dataset. First, an abundance map of the mineral was generated using the methodology proposed in section \ref{subsection: extraction of site-specific limestone endmember} utilizing the site-specific endemic signature. The generated mineral map was used to choose precursors with high and low chances for limestone availability and the distribution of the abundance values was considered to develop a criterion to select pixels and thereby determine the corresponding locations for the method validation field study. From the distribution shown in Fig. \ref{figure: abundance distribution of limestone}, the first and third quartile values were considered as the threshold values for the low and high limestone availability classes, respectively and the qualified pixels were marked on the mineral map accordingly. From this marked map, four sites were chosen as given in Fig. \ref{figure: site selection} to collect soil samples for both classes as the proposed method provides locations to survey and avoid for limestone. In a way, this is an emulation of the proposed protocol to be followed when the presence of a particular mineral (in this case limestone) has been suspected. The proposed method will return the signal to survey or not, then for the affirmative cases, the manual survey will be performed. Here through the validation field survey of the selected survey sites, the validity of using the proposed method as a precursor is tested. Then, eighty locations were chosen randomly from each site resulting in forty soil samples for each limestone availability class. After the locations were decided, in-situ data collection was performed around the designated sites. Though each pixel had 30$\times$30 m\textsuperscript{2} spatial resolution, in practice it is impossible to collect samples from such an outsized area; hence soil sample collection was executed assuming that the soil composition is homogenous. Importantly, the surface was carefully scraped to gather samples without digging into the soil, because \remotesensing~ images only correspond to material or information that is available on the surface.

\subsection{Sample preparation and laboratory testing}
\label{section: sample preparation and testing}
\begin{figure*}[h!]
    \centering
    \includegraphics[width=0.9\textwidth]{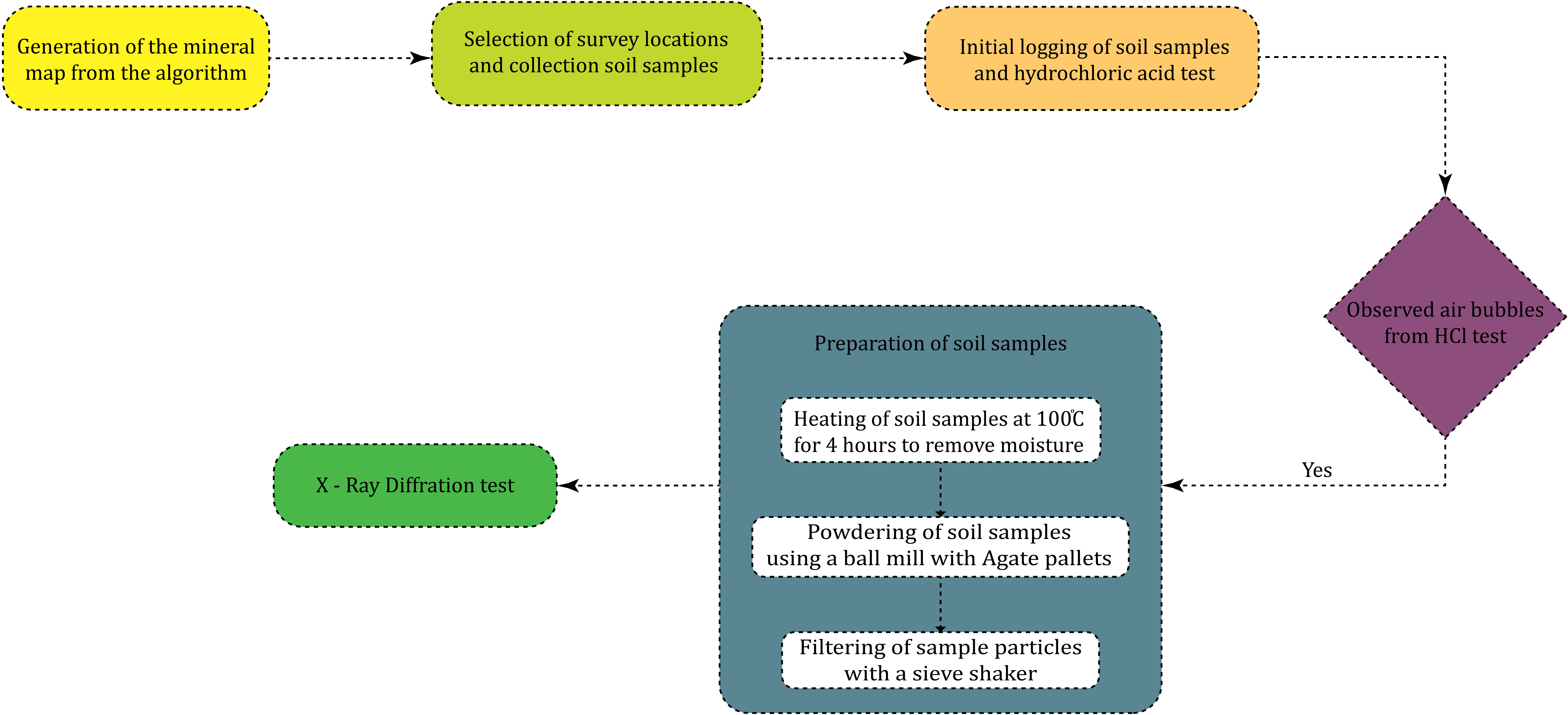}
    \caption{Flow chart of the sample preparation and laboratory tests}
    \label{figure: sample preparation flow chart}
\end{figure*}

To compare the results of the algorithm with in-situ data, the amount of limestone available in the samples should be measured. The established laboratory testing for this is the X-ray diffraction (XRD) test\citep{summit2003experiences}: a method that compares diffraction patterns of materials. Following analysis steps were followed before the sample preparation:
\begin{enumerate}
    \item The texture, color, and grain size of the samples were manually inspected and logged.
    \item Reaction with hydrochloric acid was observed as CaCO3 in limestone reacts with the acid.
    \item Samples were grouped as typical if the reaction with the acid was as expected, otherwise aberrant.
    \item Observations of the acid test were validated by checking for different compounds using a digital microscope.
    \item A total of twenty samples was selected (five samples from each site) for the XRD test and included as many mineral compounds other than limestone as possible for the sites with low abundance values.
\end{enumerate}
Then for the sample preparation:
\begin{enumerate}
    \item The sorted samples were dried at 100\,$^\circ$C for four hours using an industrial oven and were broken into pieces using a mallet.
    \item Each sample was powdered using a ball mill with 15\,mm Agate pallets under a 400 rpm setting.
    \item Each powdered sample was filtered for particles with a size smaller than 63 microns using a sieve shaker.
\end{enumerate}
Finally, the filtered samples were sent to the XRD test to find the amount of calcite, aragonite, dolomite, quartz, ilmenite, and heavy minerals. The analysis was performed using the X-ray diffractometer machine and recorded patterns from every direction of the sample by rotating the specimen holder. Fig. \ref{figure: sample preparation flow chart} is given as an illustration for the sample preparation and testing procedures.

\section{Results and discussion}
\label{section: results and discussion}

\subsection{Pre-classification of hyperspectral remote sensing data}
\label{subsection: pre-classification results}
\begin{figure*}[h!]

    \centering

    \begin{subfigure}[t]{0.23\textwidth}
        \centering
        \includegraphics[width=1.0\textwidth]{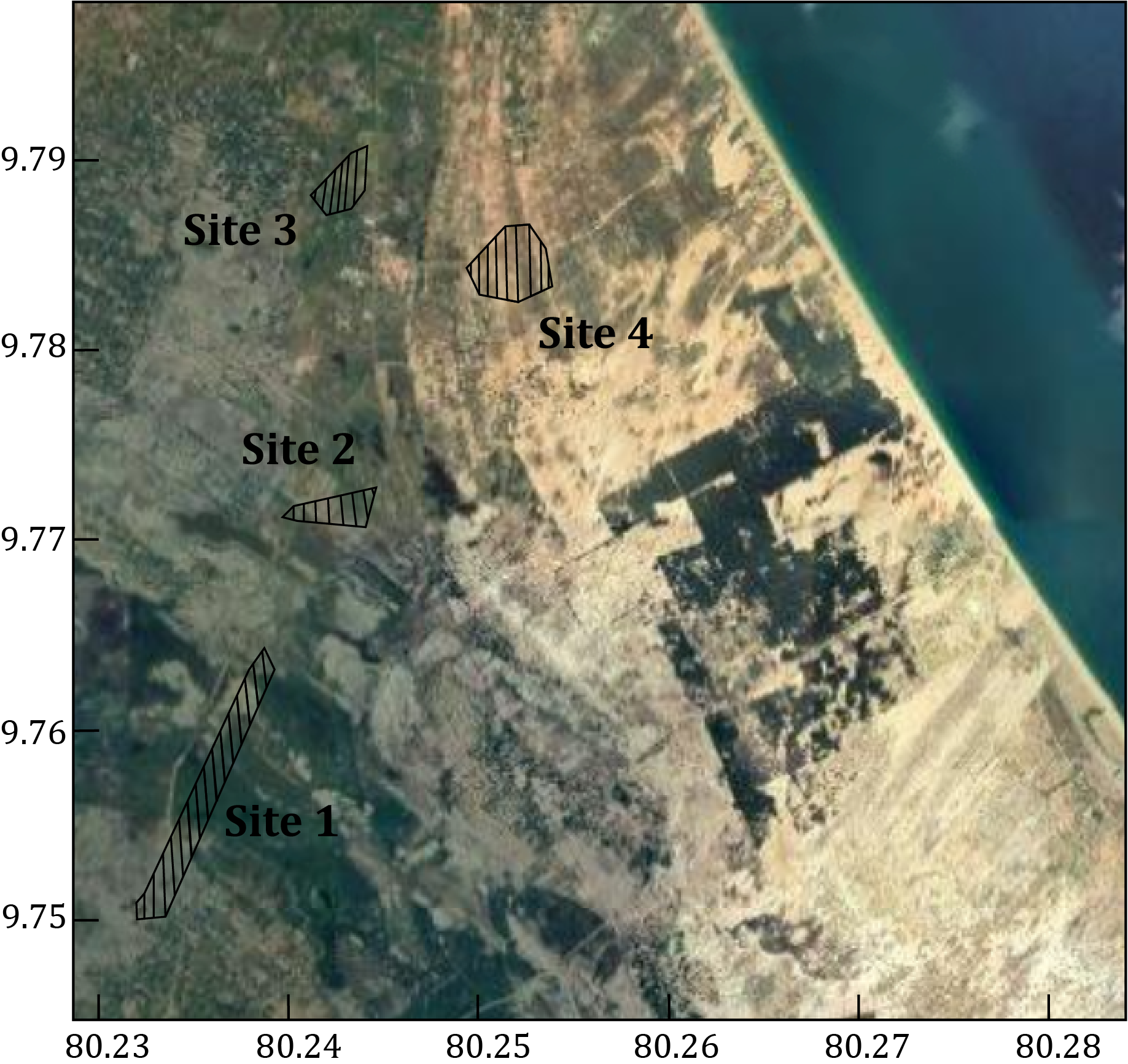}
        \caption{}
        \label{figure: true color image of location}
    \end{subfigure}
    ~
    \begin{subfigure}[t]{0.23\textwidth}
        \centering
        \includegraphics[width=1.0\textwidth]{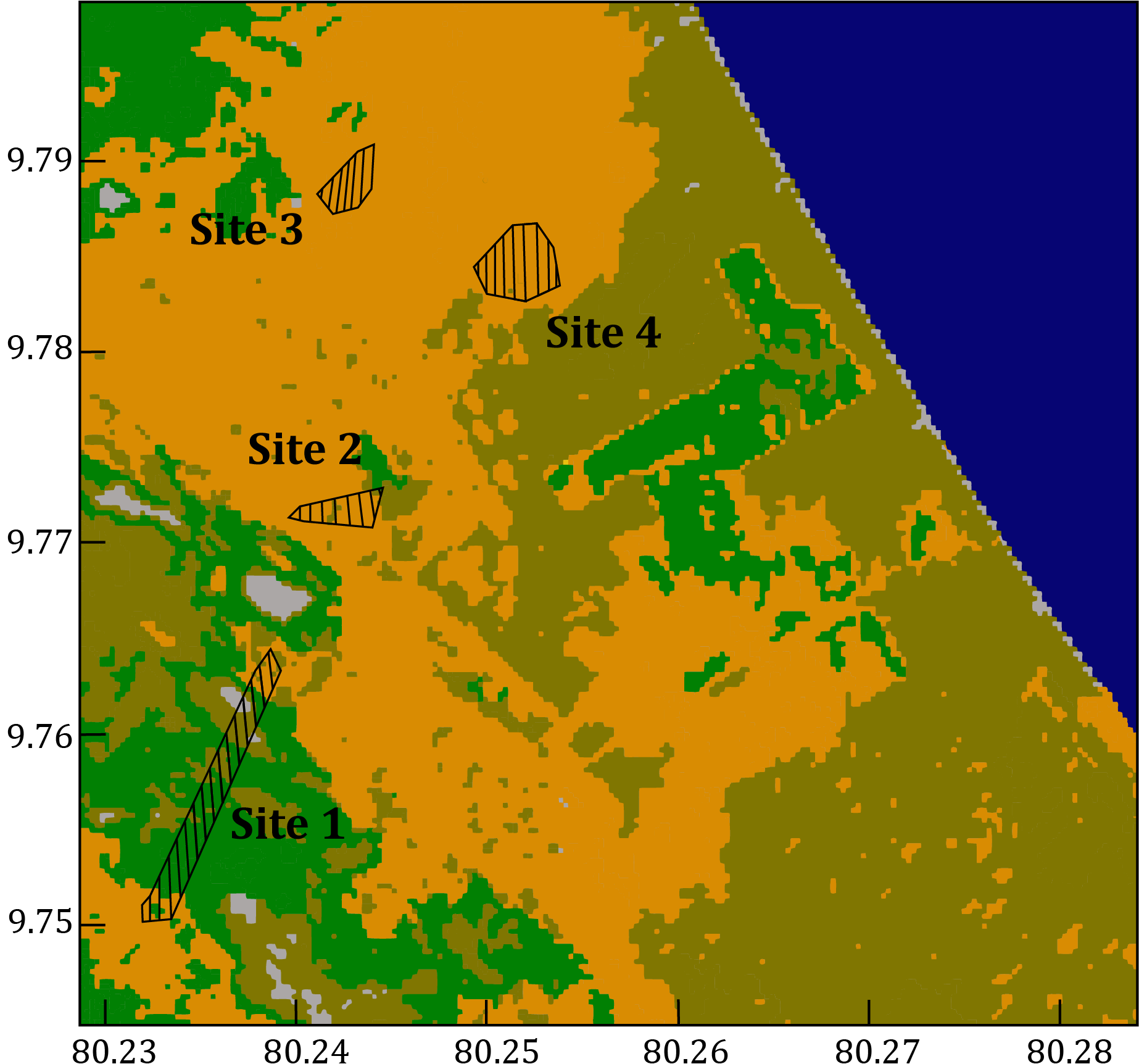}
        \caption{}
        \label{figure: sca of hyperspectral image}
    \end{subfigure}
    ~
    \begin{subfigure}[t]{0.23\textwidth}
        \centering
        \includegraphics[width=1.0\textwidth]{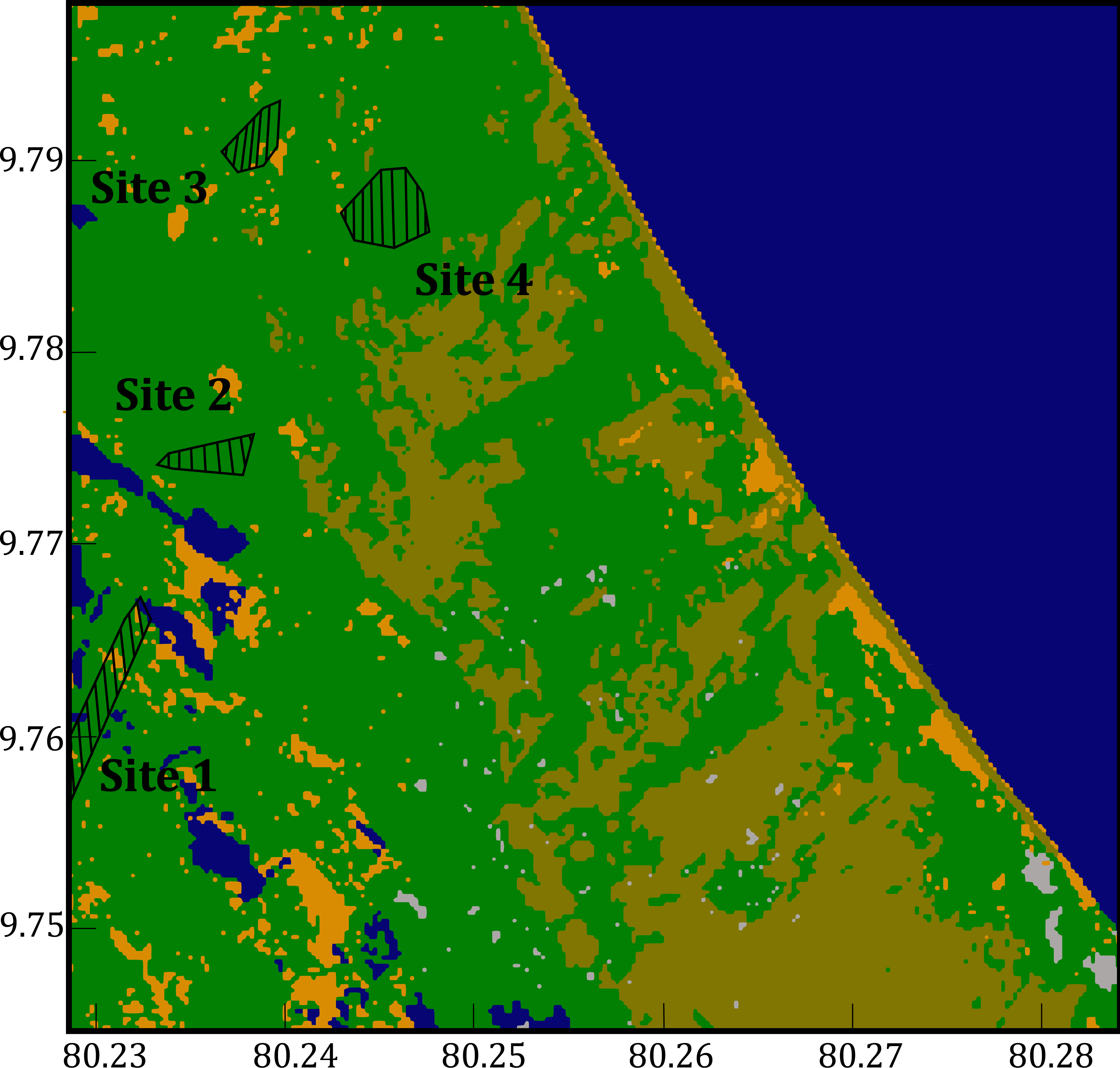}
        \caption{}
        \label{figure: vca sca multispectral}
    \end{subfigure}
    ~
    \begin{subfigure}[t]{0.23\textwidth}
        \centering
        \includegraphics[width=1.0\textwidth]{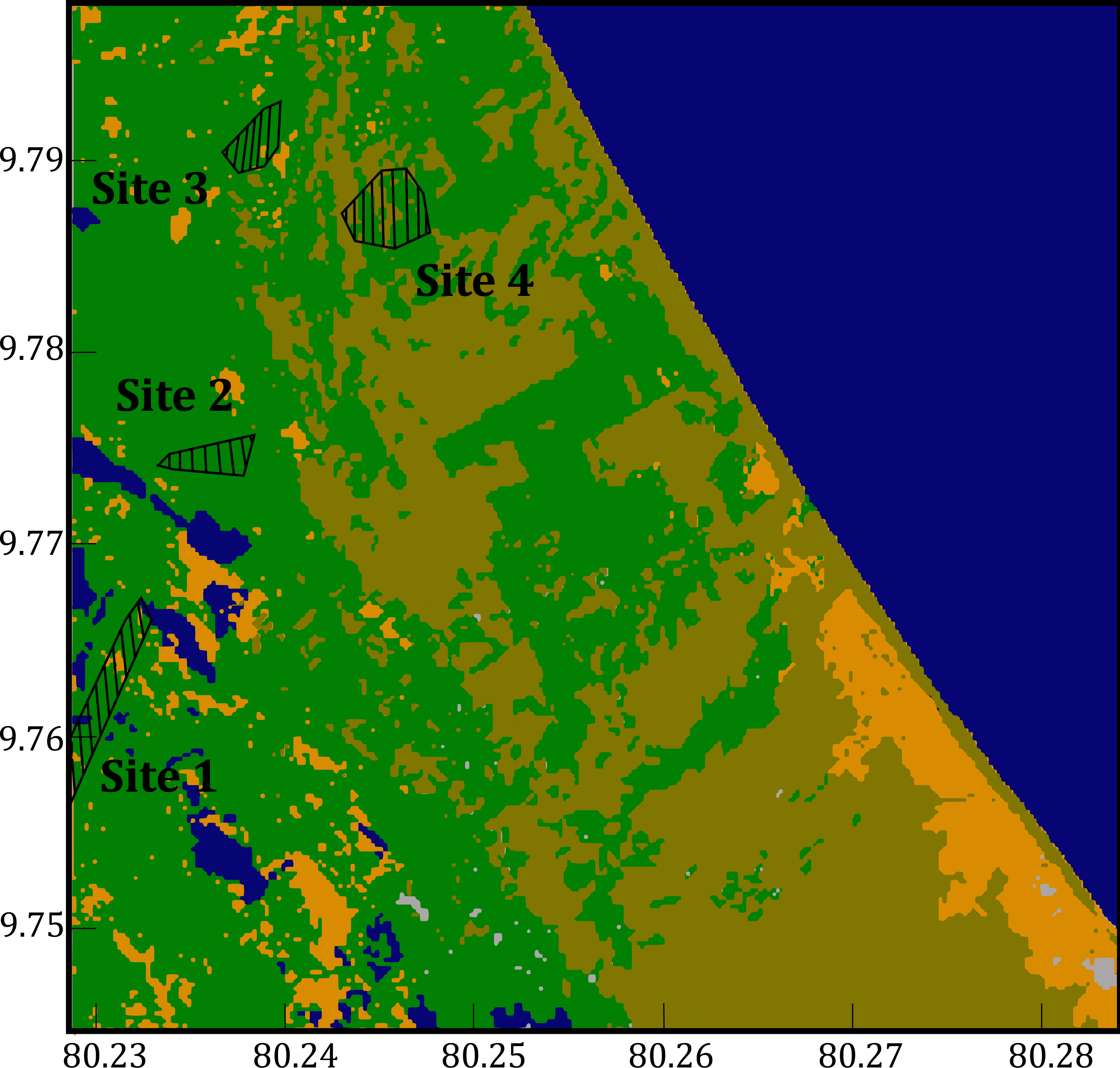}
        \caption{}
        \label{figure: autoencoder sca multispectral}
    \end{subfigure}
    
    \medskip
    
    \begin{subfigure}[t]{0.4\textwidth}
        \centering
        \includegraphics[width=1.0\textwidth]{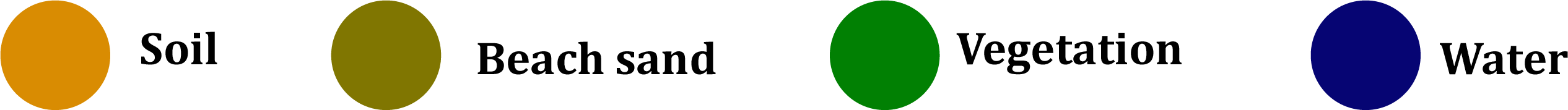}
    \end{subfigure}
    
    \caption{(a) True color satellite image with pre-classification results for (b) hyperspectral data and multispectral data under (c) \vca~and (d) autoencoder endmember extraction methods (pixels in grey color are unclassified pixels)}
    \label{figure: pre-classification results}
\end{figure*}

Pre-classification was performed to separate the soil pixels from the remotely sensed \spectralimage{H} since these soil pixels are imperative for the development of the algorithm and extraction of a close endmember.
The true-color image of the site is given in Fig. \ref{figure: true color image of location} to compare with the segmented HSI in Fig. \ref{figure: sca of hyperspectral image} under the four classes. 
The pre-classification was performed as a preparatory to limestone identification, hence in the pre-classification, pixels from other minerals were allowed into the soil class as well. This acceptance of other mineral signatures reduces the possibility to exclude pixels with a slight influence from the limestone signature.

In this work, the use of HSI was promoted as the proposed method generates a mineral map that can be used as a precursor for lithological surveys in an area polluted by non-minerals. Nonetheless, detection of an amalgam of minerals has been performed using \spectralimage{M}, especially of mineral quarries as opposed to single-target detection performed in this work. Since the mapping process itself is not a rigorous survey on the survey site, the algorithm should not profligate either computational power or financial resources to acquire remotely sensed images. Therefore, an inquiry about the most convenient kinds of inputs and a comparative analysis of their performance is eminent for the two types of most plausible inputs: \spectralimage{H}s and \spectralimage{M}s. 

To compare the effect of the input type, the pre-classification was performed with Landsat-8 multispectral data. \vca~ \citep{nascimento2005vertex} algorithm and an autoencoder \citep{hua2020autoencoder} were used for endmember extraction for pre-classification, and the segmented images are given in Fig. \ref{figure: vca sca multispectral} and \ref{figure: autoencoder sca multispectral}, respectively. When compared with classification results in Fig. \ref{figure: sca of hyperspectral image}, certain pixels that are suitable to be classified as soil pixels have been confounded as from the vegetation class as shown in Fig. \ref{figure: vca sca multispectral} and \ref{figure: autoencoder sca multispectral}, especially the pixels from the sites that contained a high percentage of limestone according to the XRD results from the field survey. Thereby, the \spectralimage{M}s have not been able to effectively resolve the constraint imposed on it due to low spectral resolution despite high noise performance simply because of failure to interpolate spectral information. Furthermore, most inorganic materials are responsive to the near and short wave infrared regions \citep{bao2016near} while most organic materials respond to wavelengths from visual and near-infrared regions \citep{heller2020effect}. Since vegetation zones and soil contain mostly organic and inorganic materials, the number of effective spectral bands that can be used to generate a signature is halved. Albeit, this is applicable for both hyperspectral and multispectral sensors, this effect exacerbates the lack of spectral information available to perform land cover classification for MSI as confirmed by the results. In addition, its performance under low SNR \spectralimage{H}s indicates that the proposed method is more noise-robust vicariously through the glut of spectral features.

\subsection{Site-specific endmember extraction}
\label{subsection: site-specific endmember extraction}

\begin{figure*}[ht!]
    \centering
    \begin{subfigure}[t]{0.45\textwidth}
        \centering
        \begin{subfigure}[t]{\textwidth}
        \centering
        \includegraphics[width=\textwidth]{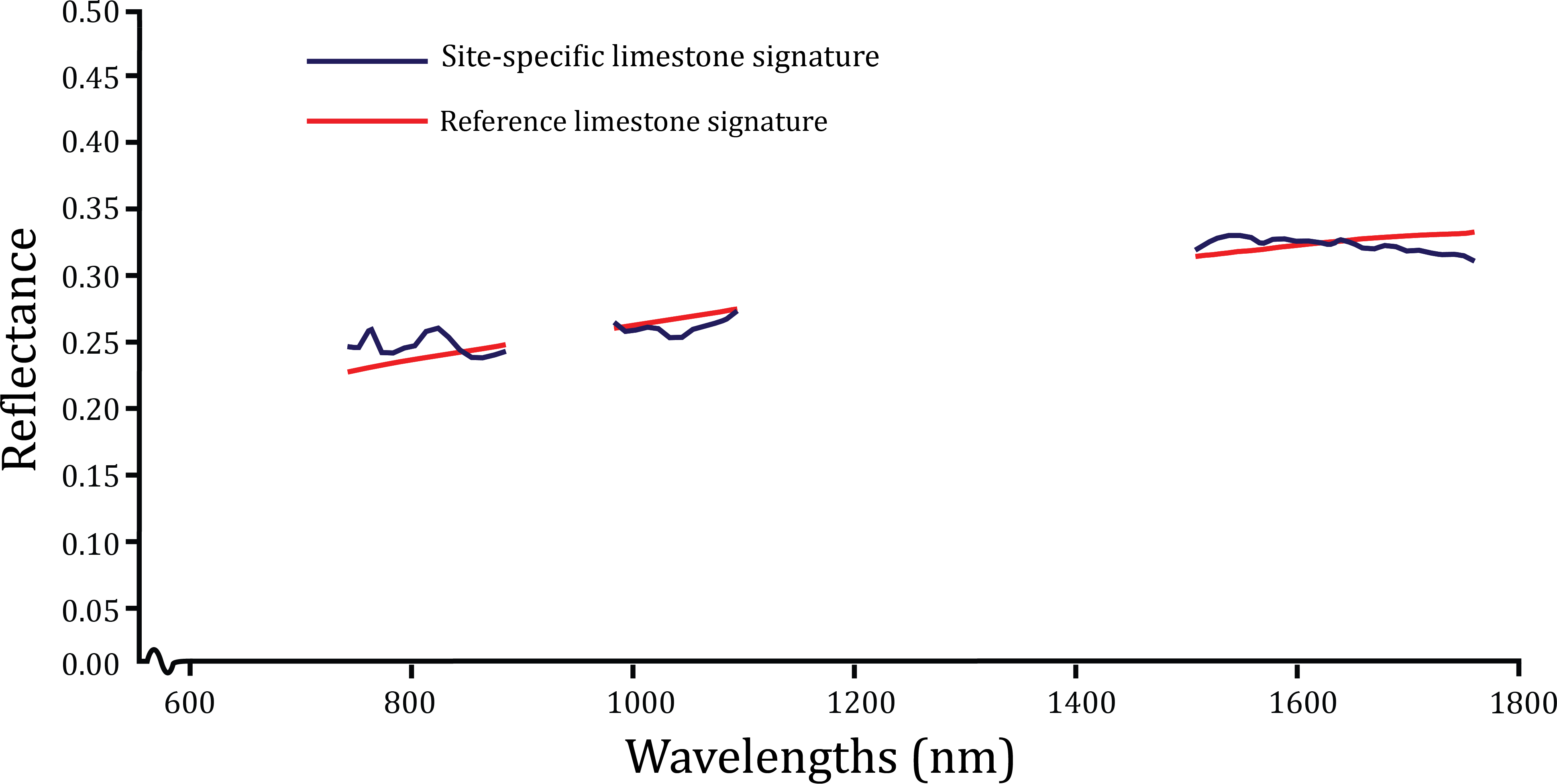}
        \caption{}
        \label{figure: site-specific endmember}
        \end{subfigure}
        
        \medskip
        
        \begin{subfigure}[t]{\textwidth}
        \centering
        \includegraphics[width=\textwidth]{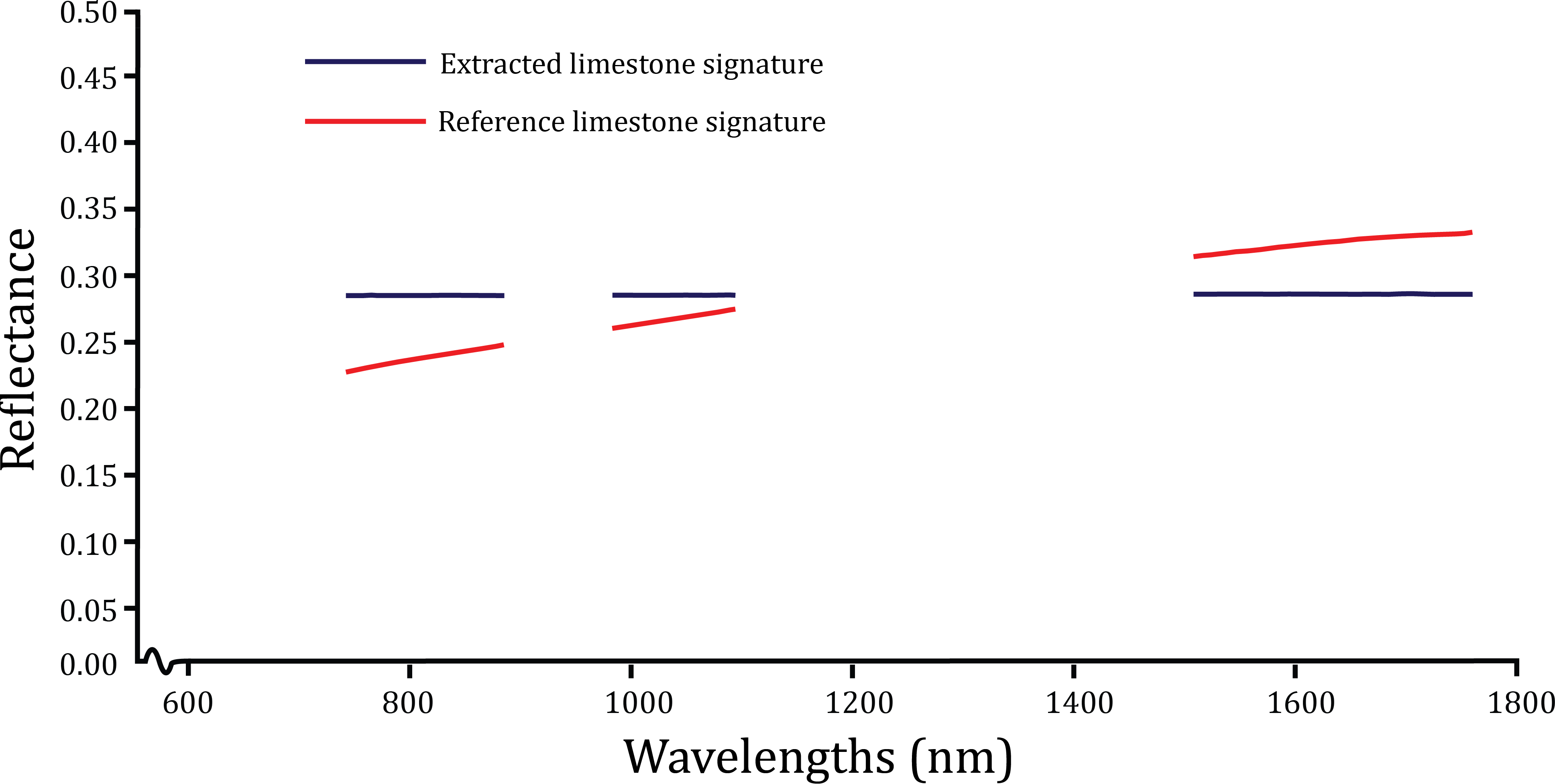}
        \caption{}
        \label{figure: window one endmember}
        \end{subfigure}
        
        \medskip
        
        \begin{subfigure}[t]{\textwidth}
        \centering
        \includegraphics[width=\textwidth]{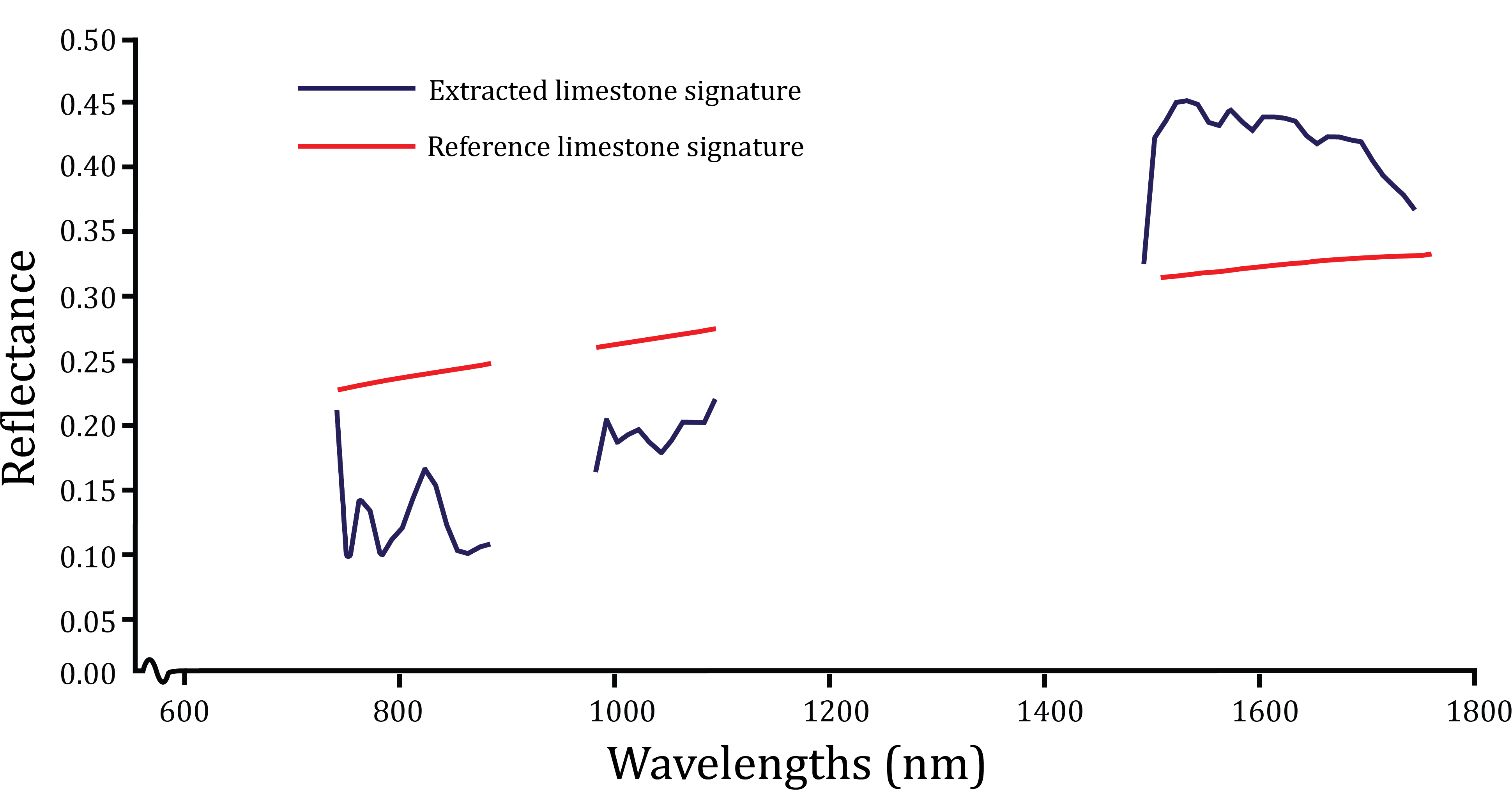}
        \caption{}
        \label{figure: window five endmember}
        \end{subfigure}
    \end{subfigure}
    ~ 
    \begin{subfigure}[t]{0.45\textwidth}
        \centering
        \begin{subfigure}[t]{\textwidth}
        \centering
        \includegraphics[width=\textwidth]{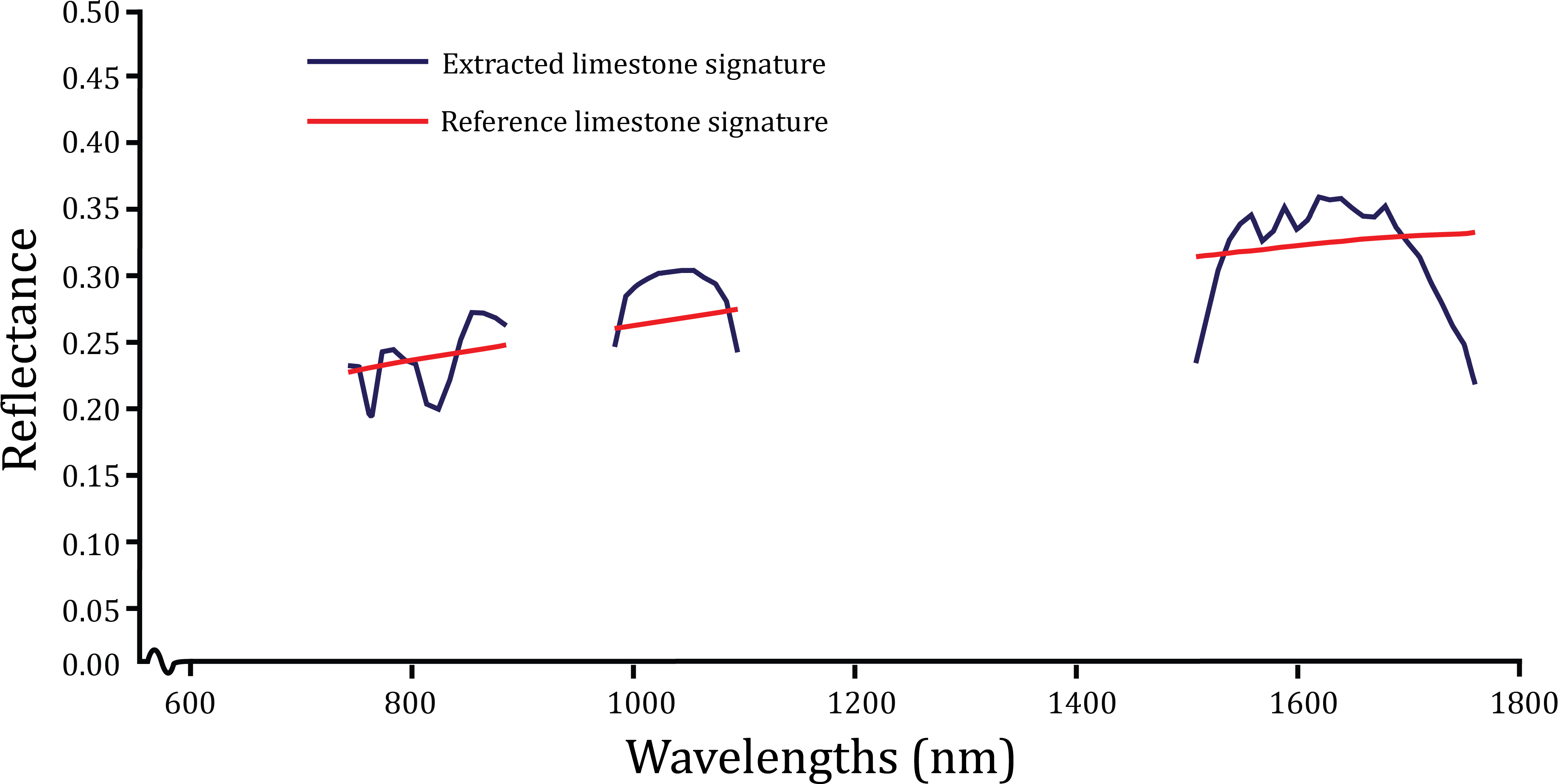}
        \caption{}
        \label{figure: extracted signature without noise removal}
        \end{subfigure}
        
        \medskip
        
        \begin{subfigure}[t]{\textwidth}
        \centering
        \includegraphics[width=\textwidth]{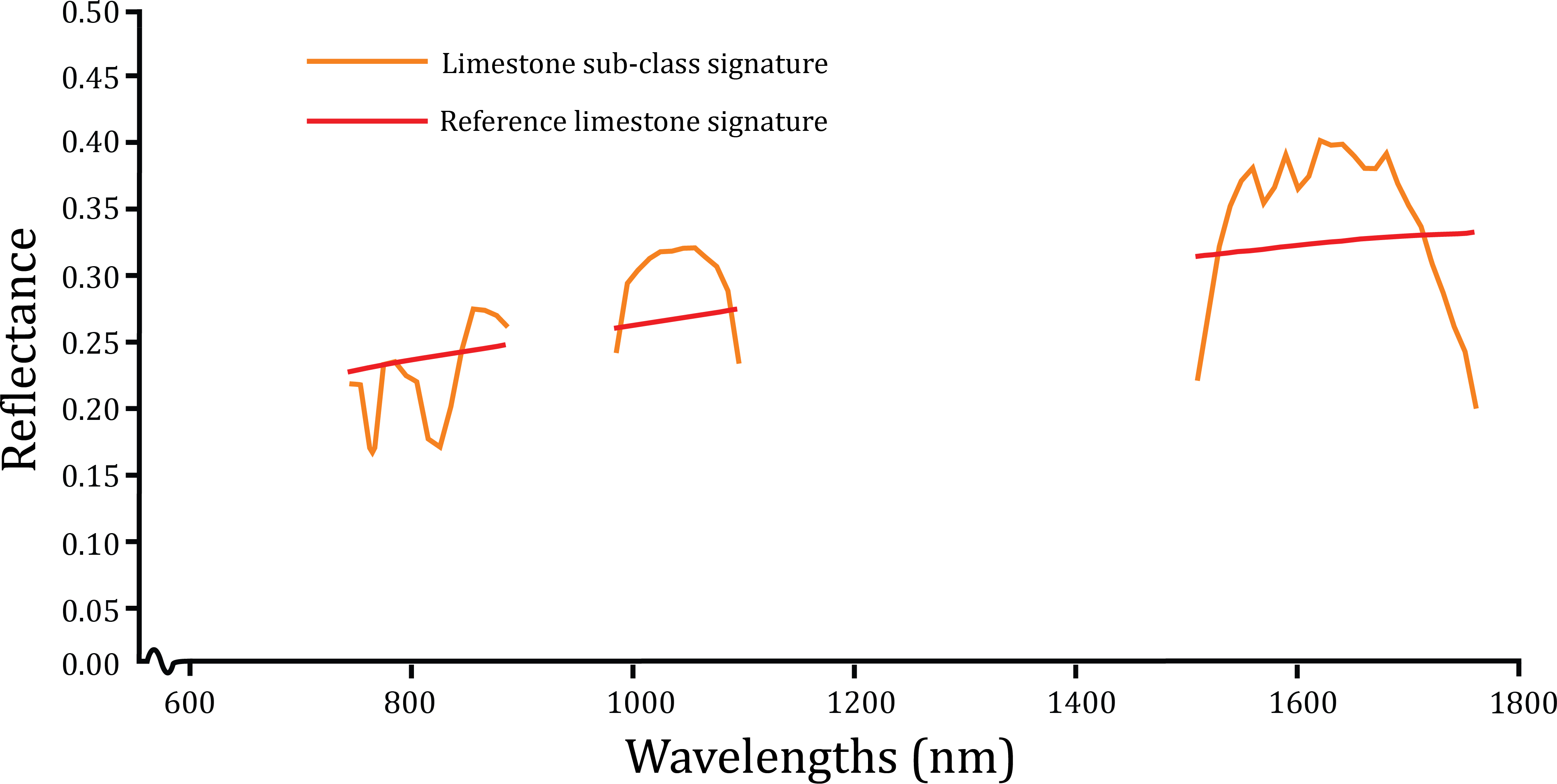}
        \caption{}
        \label{figure: limestone subclass for no noise removal}
        \end{subfigure}
        
        \medskip
        
        \begin{subfigure}[t]{\textwidth}
        \centering
        \includegraphics[width=\textwidth]{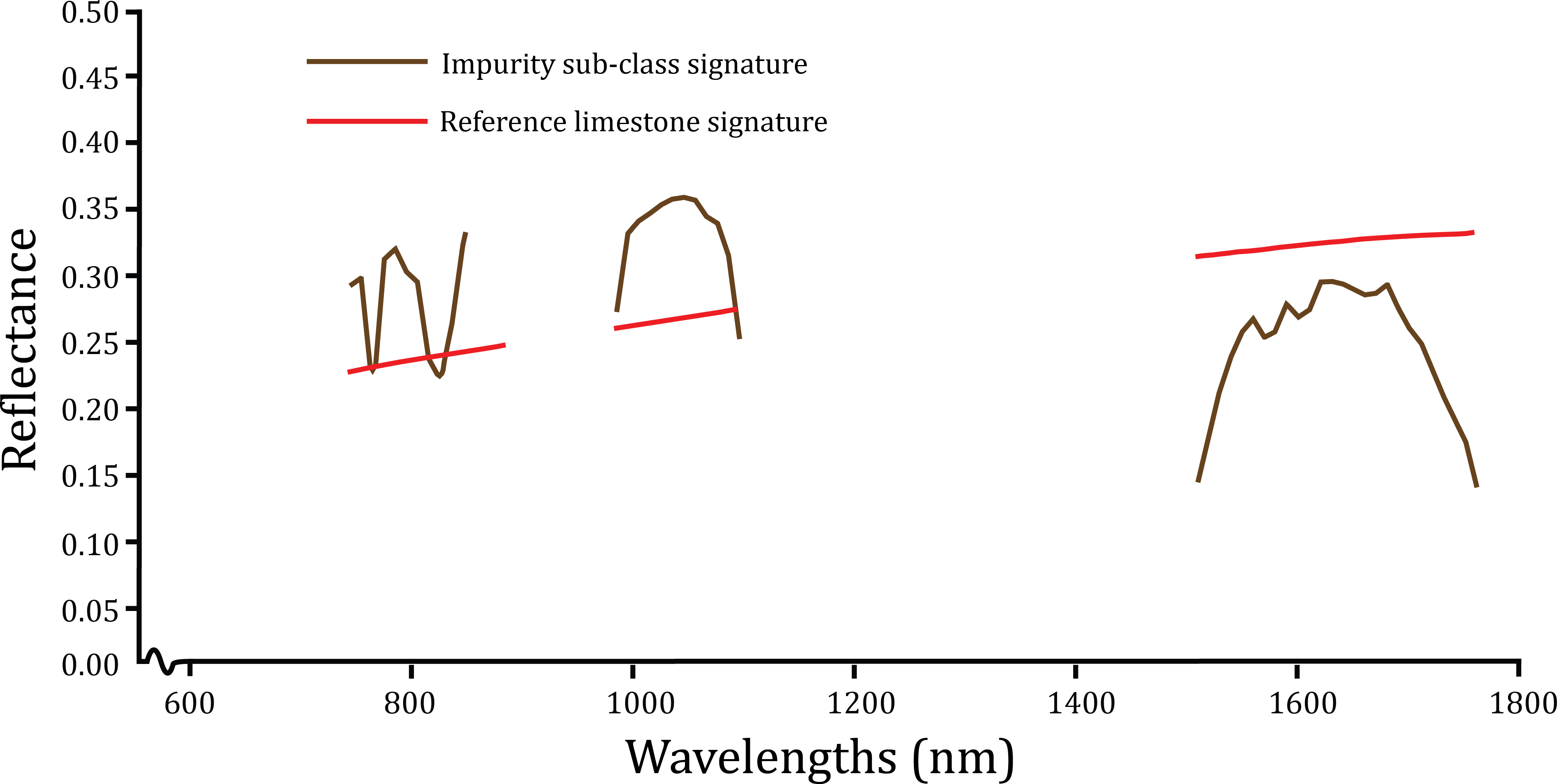}
        \caption{}
        \label{figure: impurity subclass for no noise removal}
        \end{subfigure}
    \end{subfigure}

    \caption{Results for the site-specific endmember extraction, effect of the filter window length, and effect of the residual impurities signature removal (Conditions and remarks of spectral signatures are provided in Table \ref{table: conditions and remarks})}
    \label{figure: endmember extraction results}
\end{figure*}

The objective of the endmember extraction in the algorithm was to generate a limestone signature that is endemic to the survey region, yet inherits the spectral characteristics of a pure limestone obtained by laboratory spectroscopy as such signature is a better representative for the mineral signature in the presence of other constituents but not necessarily minerals. Generally, endmember extraction is performed as a routine exercise of \remotesensing~ data unmixing with no prior knowledge about the endmembers except for the number of endmembers which is not a prerequisite with the single-target mineral detection. Also, the constraint optimization in unmixing is imposed globally (i.e. the consideration of all the endmembers in the image is a necessity), hence, might not be a feasible strategy for perfect mineral signature detection in a site-specific manner in the presence of residual impurities as done so by the proposed methodology. Therefore, a Wiener filter in the inverse modeling arrangement as given in Fig. \ref{figure: wiener filter arrangement} was used to extract a limestone endmember from the hyperspectral dataset remotely. To select the optimal Wiener filter setting, endmembers extraction was performed for different filter lengths and the setting with the lowest RMSE was considered as the optimal setting. The optimal error metric was recorded for filter length, thereby the filter setting was set accordingly. The output of the Wiener filter (extracted endmember) is shown in Fig. \ref{figure: site-specific endmember} along with the laboratory reference signature, and it is conspicuous from Fig. \ref{figure: site-specific endmember} that the extracted endmember is an approximate representation of the laboratory reference but encapsulates a majority of the spectral variations from the pixels that are greatly correlated with the pure limestone signature. Since the input signals to the Wiener filter should be zero mean, the output of the filter or the extracted endmember was shifted using the mean value of the laboratory reference signature and an RMSE, and a correlation coefficient of $0.0754$ and $0.9639$ were recorded between the site-specific signature and the library reference signature.

\begin{table*}[ht!]
\centering
\captionsetup{labelsep=newline, width=80mm}
\caption{Conditions and remarks of spectral signatures of Fig. \ref{figure: endmember extraction results}}
\label{table: conditions and remarks}
\begin{tabular}{>{\centering}p{5mm}>{\centering}p{30mm}>{\centering}p{30mm}>{\arraybackslash}p{80mm}}
\hline\hline
Index & Filter window length & Removal of the residue signature & Remarks\\
\hline
(a) & 3 & applied & site-specific endmember for limestone extracted at the filter output\\
(b) & 1 & applied & extracted endmember at the filter output assuming no correlation amongst adjacent spectral bands\\
(c) & 5 & applied & extracted endmember at the filter output assuming strong correlation amongst spectral bands over a larger interval\\
(d) & 3 & not applied & extracted endmember at the filter output without removing the residual impurities signature from the input to the filter\\
(e) & - & not applied & limestone subclass signature used to extract the site-specific signature\\
(f) & - & - & residual impurities subclass signature used to modify the input to the filter\\
\hline\hline
\end{tabular}
\end{table*}

To examine the effect on the site-specific signature from the filter size, the resulting endmembers were considered as the filter size was varied and compared against the optimal result (Fig. \ref{figure: site-specific endmember}). When the filter size decreased to one, the resulting endmember (Fig. \ref{figure: window one endmember}) was approximated to a straight line; while the filter size increased, the extracted signature contained aberrant fluctuations. In the first case, a first-order filter is suitable for the unmixing problem, only if the impurity signature used for signal conditioning was ideal because then the filter coefficient is tantamount to the amount of limestone available in the pixel. But the signatures used in the algorithm are neither pure nor noise-free, therefore the Wiener filter can only perform an optimal scaling on the input signal. In the second case, with a filter order of five, the spectral bands are assumed to be largely correlated and when compared with the laboratory reference signature, correlation amongst consecutive bands should be positive. Nonetheless, it is visible from the spectral signature in Fig. \ref{figure: limestone subclass for no noise removal} of the limestone subclass that certain spectral bands are negatively correlated (i.e. the trend in nearby spectral bands are opposite). This mismatch has affected the output of the Wiener filter because according to the principle of correlation cancellation, the filter only attempts to reconstruct the correlated portion of the input signal with the desired signal and therefore has introduced undesired fluctuations (Fig. \ref{figure: window five endmember}) at the output.

The effect of not removing the estimated impurity signature is illustrated in Fig. \ref{figure: extracted signature without noise removal} for the optimal filter length of three. According to Fig. \ref{figure: extracted signature without noise removal}, it is noticeable when the input to the Wiener filter was not conditioned by removing the impurity signature, the output of the filter, i.e. the extracted endmember for limestone is not an approximation for the desired or laboratory reference signal, but is a combination of the mean signature of the high and low limestone available pixels. The reason is, though the limestone subclass pixels (Fig. \ref{figure: limestone subclass for no noise removal}) are assumed to be pure pixels, these still could contain reflectance properties from other mineral classes. Hence, if the desired signal correlates with the signatures from these other classes (Fig. \ref{figure: impurity subclass for no noise removal}), it is guaranteed that the desired signal will be highly correlated with the low limestone available signatures as well. Even though theoretically, these signatures are derived from independent and uncorrelated sources, practically these sources will be correlated as minerals can have similar responses at least at certain spectral bands due to their chemistry. Hence, if the impurity signature was not removed from the soil pixels used for the extraction, the input signature to the filter might be an amalgam of multiple compounds. Thereby, the basic correlation matching will not be as efficacious as if there were only two components: one from a correlated source and the other from an uncorrelated source. So, removing the effect of the correlated residual impurities using a site-specific estimated residual impurity signature, though might not emulate the signatures of every mineral in the region, will minimize the possibility of corruptive effects of those undesired signatures to contain correlated components on the site-specific endemic signature generation process.

\subsection{Generated mineral maps for predictive modeling}
\label{subsection: generated mineral maps}

\begin{figure*}[h!]
    \centering
    
    \begin{subfigure}[t]{0.32\textwidth}
    \centering
    \includegraphics[width=\textwidth]{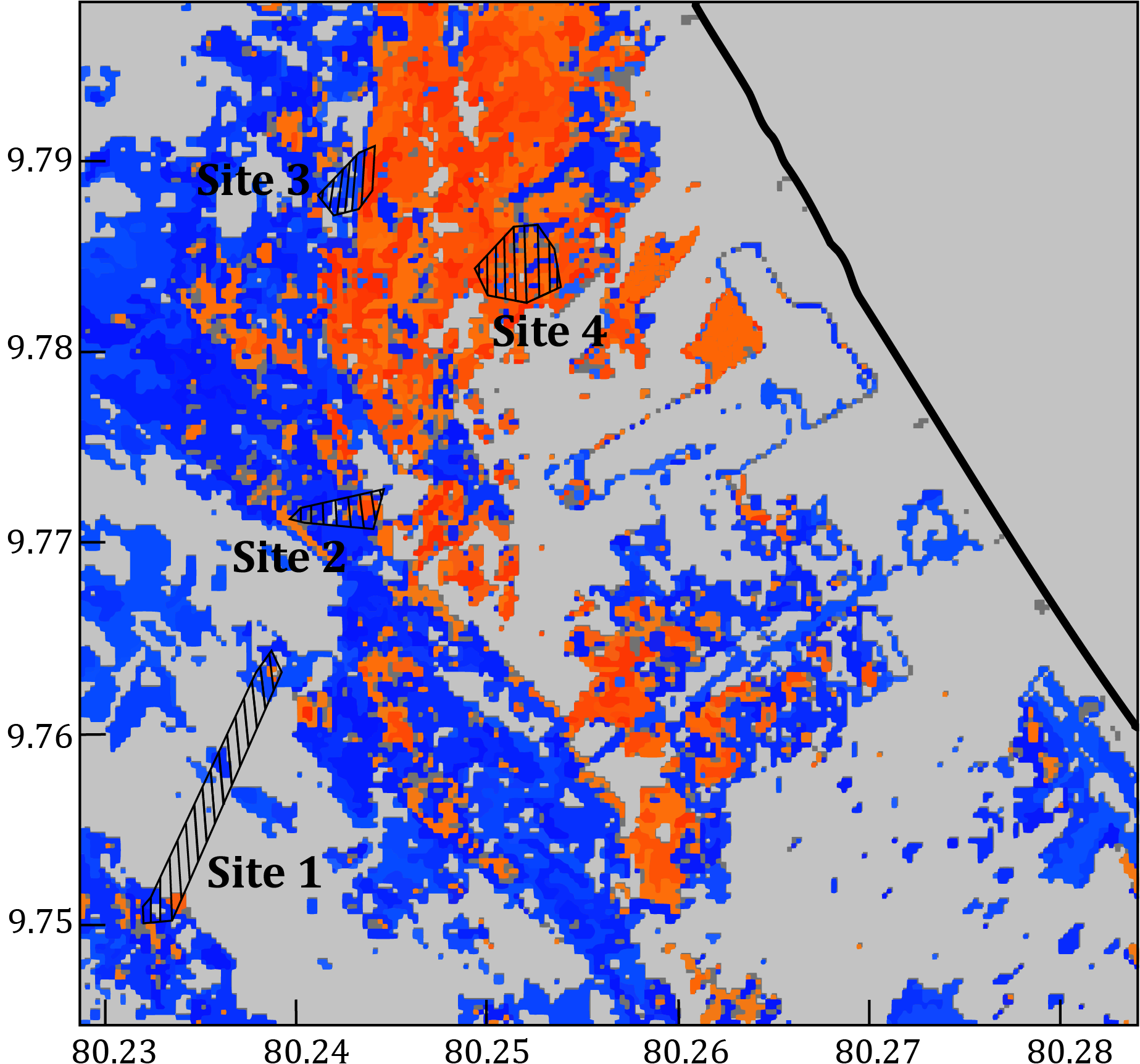}
    \caption{}
    \label{figure: generated mineral map}
    \end{subfigure}
    \,
    \begin{subfigure}[t]{0.32\textwidth}
    \centering
    \includegraphics[width=\textwidth]{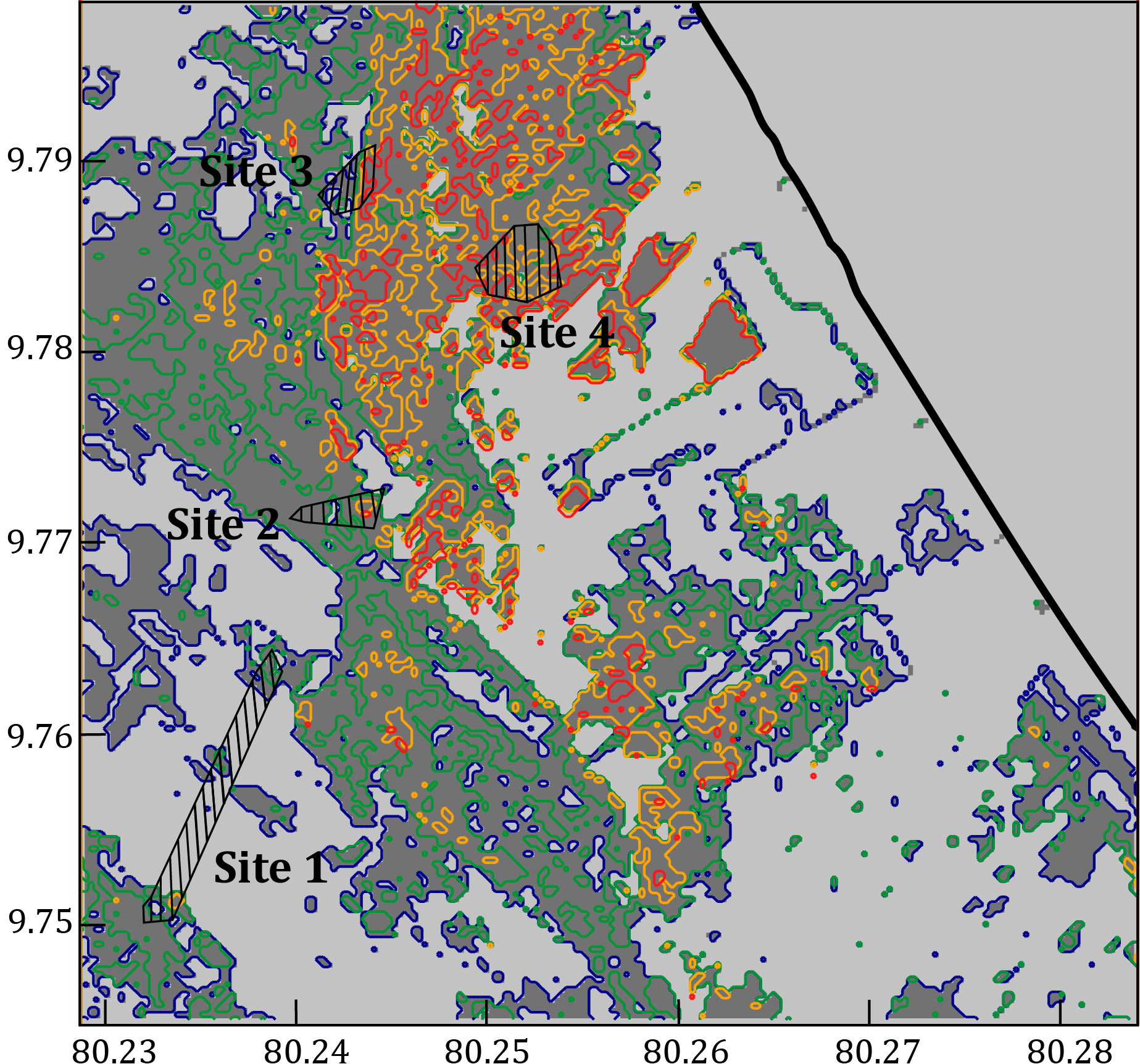}
    \caption{}
    \label{figure: generated contour map}
    \end{subfigure}
    \, 
    \begin{subfigure}[t]{0.32\textwidth}
    \centering
    \includegraphics[width=\textwidth]{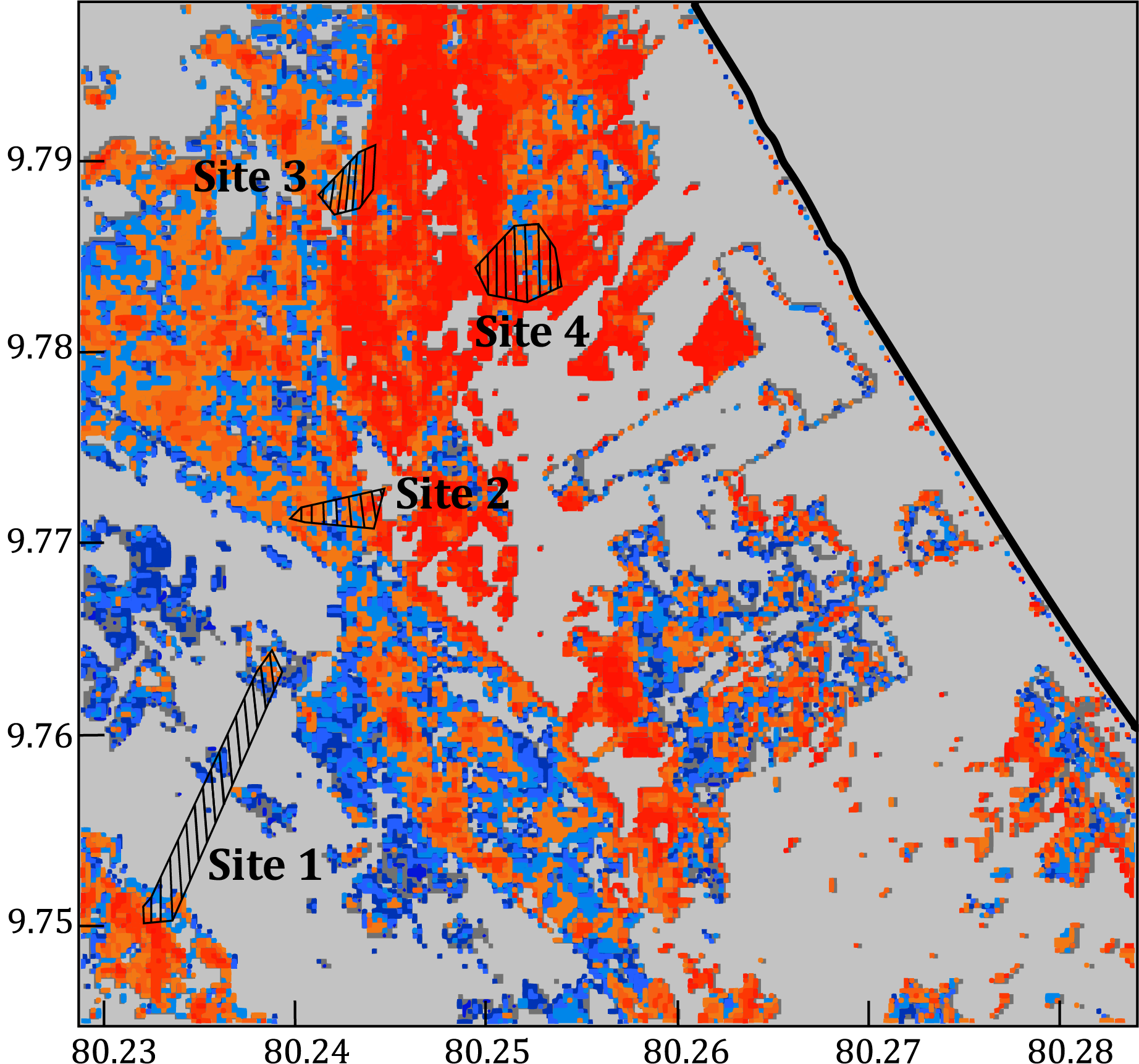}
    \caption{}
    \label{figure: mineral map without noise removal}
    \end{subfigure}
    
    \medskip

    \begin{subfigure}[t]{0.40\textwidth}
    \centering
    \includegraphics[width=\textwidth]{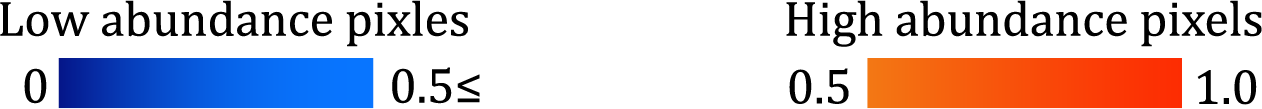}
    \caption*{Legend for the mineral maps (a) and (c)}
    \end{subfigure}
    \hspace{10mm}
    \begin{subfigure}[t]{0.40\textwidth}
    \centering
    \includegraphics[width=\textwidth]{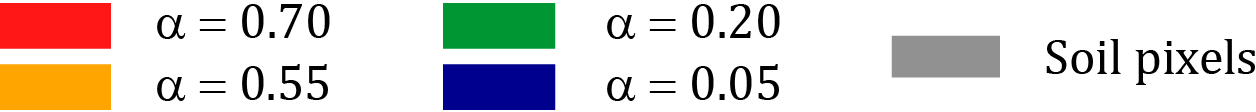}
    \caption*{Legend for the contour map (b)}
    \end{subfigure}

    \caption{Generated (a) mineral map and (b) contour map with stochastic cancellation and residual impurity removal, and (c) the mineral map without stochastic cancellation and residual impurities removal}
    \label{figure: mineral and contour maps}
\end{figure*}

The derived mineral map for limestone given in Fig. \ref{figure: generated mineral map} was constructed using the abundance values ($\alpha$) calculated according to \ref{equation: limestone abundance estimator} with the site-specific limestone signature, for each soil pixel. The site-specific limestone signature used for abundance calculation was extracted with a filter order of three as delineated in section \ref{subsection: site-specific endmember extraction} and is illustrated in Fig. \ref{figure: site-specific endmember}. When observing the resulted $\alpha$ values, certain pixels contained negative values; hence removed from the map as these pixels are less likely to contain limestone. These aberrations must have happened because of, the unmixing rule given in \ref{equation: limestone abundance estimator} uses the properties of the linear mixing model and an estimated signature for residual impurities', and assumes noise-free spectral signatures. The generated mineral map has a smooth and continuous variation in the availability of limestone which agrees with the natural variation in the soil composition. Besides, the contour map shown in Fig. \ref{figure: generated contour map} generated for the mineral map describes the overall variation in the soil composition of limestone. According to Fig. \ref{figure: generated contour map}, the closed-loop contour lines denote the change in limestone composition in the soil is changing gradually and concentrated to certain regions; and this variation of the contours is similar to that of a peninsula of pure minerals. Though, the algorithm was not constrained to promote the smoothness in the abundances for limestone the resulting mineral map has inherited the said property with the proposed method.

\begin{table*}[h!]
\centering
\captionsetup{labelsep=newline, width=140mm}
\caption{Mean abundance parameter values of sites under the proposed method (with stochastic cancellation and residual impurity removal), without the proposed signature cancellation, and XRD test}
\label{table: mean limestone parameter values}
\begin{tabular}{p{40mm}>{\centering}p{20mm}>{\centering}p{20mm}>{\centering}p{20mm}>{\centering\arraybackslash}p{20mm}}
\hline\hline
Parameter & Site-1 & Site-2 & Site-3 & Site-4\\
\hline
$\alpha$ with site-specific endmember-proposed & 0.083 & 0.2022 & 0.0572 & 0.8034\\
$\alpha$ without stochastic cancellation & 0.547 & 0.583 & 0.775 & 0.872\\
XRD percentages & 1.418 & 24.244 & 4.882 & 48.866\\
\hline\hline
\end{tabular}
\end{table*}

The method proposed in the work used a single-target identification approach to map the abundances of limestone, hence the algorithm is amenable to consider minerals other than limestone and overestimate the abundance of the target mineral. To reduce the inclusive nature of the method, we are using stochastic signal processing to extract a spectral signature that has characteristics similar to laboratory reference but has been adjusted for spectral variations indigenous to the pixels from the study area. The results in Fig. \ref{figure: generated mineral map} and \ref{figure: generated contour map} utilized this stochastic cancellation mechanic to generate the mineral map and contour plot. In addition, the mineral map generated with the limestone subclass signature (without stochastic cancellation and residual impurity removal) is given in Fig. \ref{figure: mineral map without noise removal}. According to Fig. \ref{figure: mineral map without noise removal}, the estimated abundances without stochastic cancellation are higher than that of with stochastic cancellation as tabulated in Table \ref{table: mean limestone parameter values}. 
To compare the improvement on the abundance calculation from the use of stochastic cancellation and residual impurity removal, the $\alpha$ values from the two methods were compared with a laboratory experiment known as XRD test, which was performed to analyze the composition as given in Table \ref{table: mean limestone parameter values}. When compared to the XRD results, the $\alpha$ values calculated without stochastic cancellation and residual impurity removal are severely higher than that of the proposed method. In the XRD test, the diffraction pattern of the specimen is compared with known patterns of minerals to establish the composition of the sample. It should be noted that the abundance values range from 0-1 while the XRD results are given in percentages.

\begin{figure}[h!]
    \centering
    \includegraphics[width=\columnwidth]{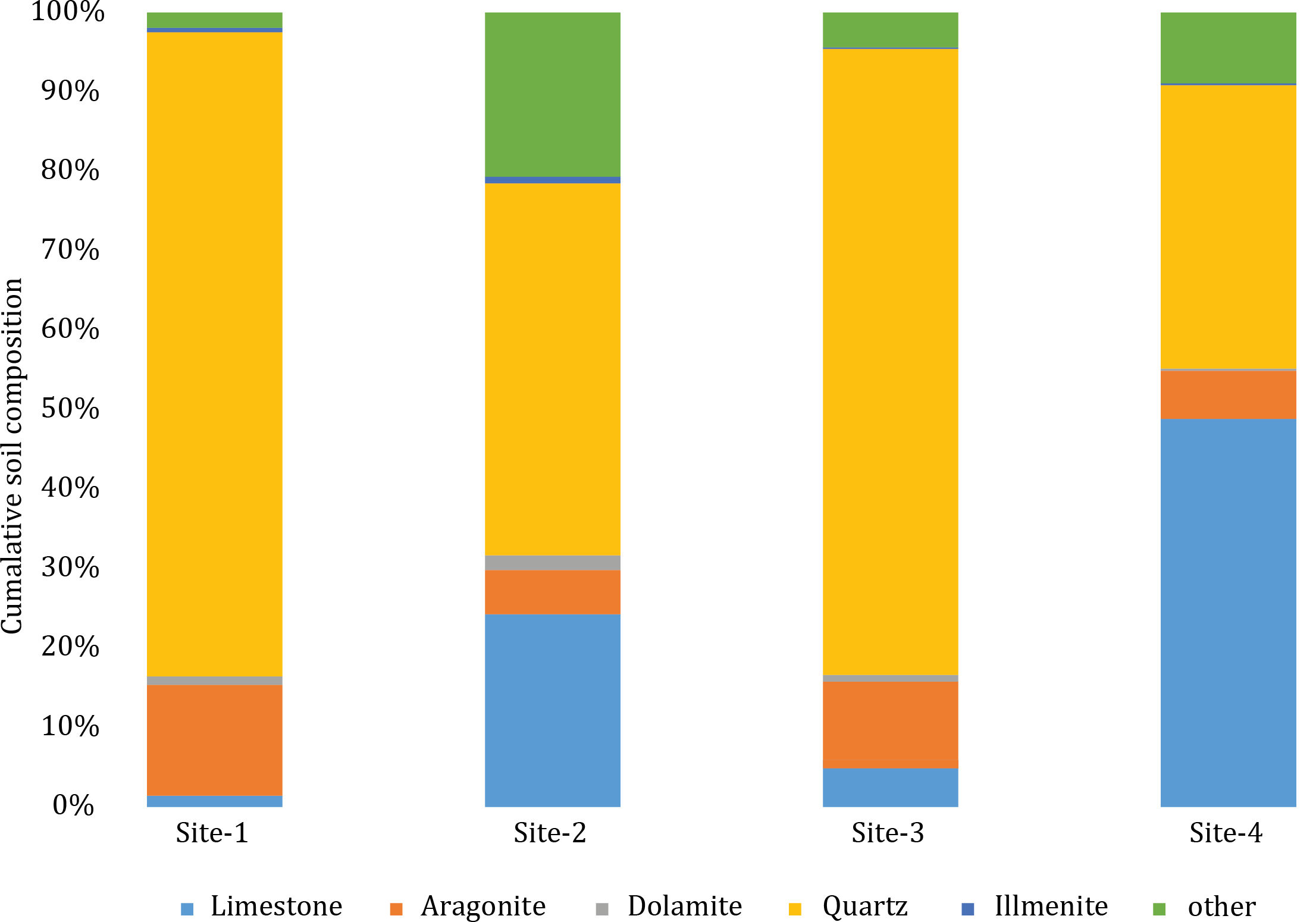}
    \caption{Composition of soil from field survey with selected constituents}
    \label{figure: soil composition}
\end{figure}

This overestimation of limestone abundance is possibly due to the spectral similarity with the limestone reference signature of other mineral constituents. Thereby, the use of stochastic cancellation has ameliorated the differentiation between the signatures of the limestone and other constituents. Further, with the Wiener filter arrangement, the extracted endmember is uncorrelated from the signatures from the residual impurities. This is equivalent to using laboratory references of other possible minerals, unless the laboratory references of those minerals are correlated with the target mineral signature, in which case the inclusive nature is extant in the result. To signify the performance of the algorithm in estimating abundances under single-target identification in the presence of other minerals and non-minerals, the soil composition of each site as examined from the XRD test is illustrated in Fig. \ref{figure: soil composition}.

The stochastic cancellation-based site-specific endmember extraction was performed to facilitate single-target identification of limestone contrary to detecting a collection of minerals. But to determine the minerals captured in a pixel, we have to either have information or make assumptions on the number of principal constituents available. The former method requests for an exhaustive search on the survey area which is difficult to conduct, and the latter method would be futile unless the principal minerals are dominant. Whereas, only a reference signature for the target mineral is required to estimate the abundances in the proposed work. Besides that, the collected samples from the field survey contained other minerals as well. However, the likelihood of limestone from these samples still held as evidenced from the XRD test results in the presence of other minerals. Further, as and when new information about minerals are found from the survey the proposed method is applicable for specific minerals under the single-target approach. Though the proposed method is built upon the limestone mineral, the algorithm is generic and makes no specific assumptions for limestone. Hence, the proposed work is scalable and should apply to other minerals as well.

\subsection{Efficacy of the proposed method}
\label{subsection: xrd test of samples}
\begin{figure*}[h!]
    \centering
    
    \begin{subfigure}[t]{0.2\textwidth}
        \centering
        \includegraphics[width=\textwidth]{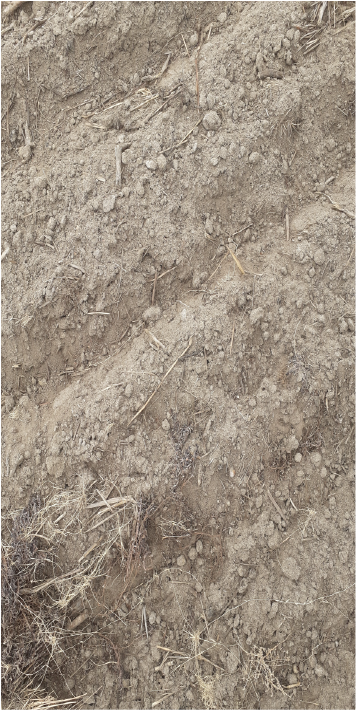}
        \caption{}
        \label{figure: site-1}
    \end{subfigure}
    ~
    \begin{subfigure}[t]{0.2\textwidth}
        \centering
        \includegraphics[width=\textwidth]{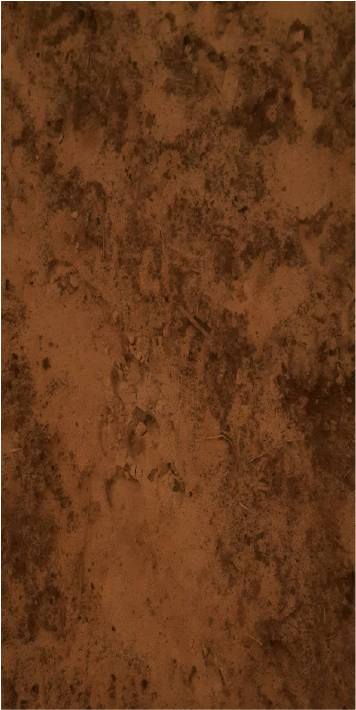}
        \caption{}
        \label{figure: site-3}
    \end{subfigure}
    ~
    \begin{subfigure}[t]{0.2\textwidth}
        \centering
        \includegraphics[width=\textwidth]{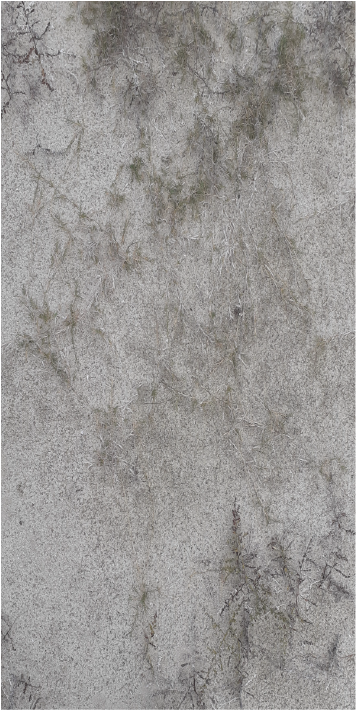}
        \caption{}
        \label{figure: site-2}
    \end{subfigure}
    ~
    \begin{subfigure}[t]{0.2\textwidth}
        \centering
        \includegraphics[width=\textwidth]{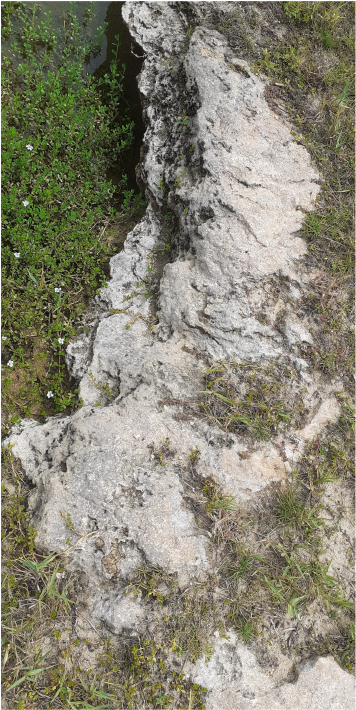}
        \caption{}
        \label{figure: site-4}
    \end{subfigure}

    \caption{In-situ soil charateristic observations for (a) site-1, (b) site-3, (c) site-2, and (d) site-4}
    \label{figure: soil characteristic visuals}
\end{figure*}

The soil characteristics and surface composition of the four survey sites are illustrated in Table \ref{table: soil characteristics} and given in Fig. \ref{figure: soil characteristic visuals}, respectively. Site-1 (Fig. \ref{figure: site-1}) was a sparse vegetation land with a dry and solid soil texture, and evidence for limestone was not available around the site. Similarly, site-3 (Fig. \ref{figure: site-2}) had no indication of surface limestone and the reddish-brown soil color was different from the other three sites. However, the area included in site-3 was an inhabited region and has had continuous interventions by the locals; hence the soil composition may have changed abruptly as opposed to the expected composition from paleogeography. On the other hand, site-2 (Fig. \ref{figure: site-3}) had moist dark brown soil and contained fragments of seashells on the surface of the region. The area is considered to be a drained lagoon and is large enough to be sensed via satellite images. There was an even spread of shell residues throughout the lagoon area and these shells should have appeared as a result of tides, and the existence of seashells is evidence for the existence of limestone on the surface. Finally, site-4 had a desiccated soil texture and had shell fragments as evidence for soil limestone. Besides, medium size limestone fragments were found around site-4 during the survey. The selected location was the closest to the shore and these sediments may have been deposited during the development of the Jaffna peninsula with the aid of the sea currents. As tabulated in Table \ref{table: mean limestone parameter values} the proposed method had made an accurate prediction of the limestone presence in site-2 and site-4 with higher $\alpha$ values while returning low $\alpha$ values for site-1 and site-3 that implies less chance of limestone presence, which is in agreement with the observations made by the validation field survey via XRD tests.

\begin{table*}[h!]
\centering
\captionsetup{labelsep=newline, width=100mm}
\caption{Soil color, texture, and characteristics of the site locations}
\label{table: soil characteristics}
\begin{tabular}{p{20mm}p{30mm}p{30mm}>{\arraybackslash}p{60mm}}
\hline\hline
\pbox{15cm}{\medskip Site\\Location\medskip}	&Soil color	&Soil texture	&Remarks\\
\hline
site-1 &Light mud-brown &Dry and dusty soil	& \pbox{60cm}{No visual evidence for limestone \\ Area is a sparse vegetation field}\medskip\\
site-2 &Dark mud-brown &Moist soil	& Fragments of sea-shells were found\medskip\\
site-3 &Reddish brown &Dry and hard soil	& \pbox{60cm}{No visual evidence for limestone \\ Often interrupted by the inhabitants}\medskip\\
site-4 &Light brown &Dry and hard soil	& \pbox{60cm}{Existence of fragments of sea-shells \\ Existence of medium size fragments of\\ sedimentary limestone}\medskip\\
\hline\hline
\end{tabular}
\end{table*}

\begin{figure*}[h!]
    \centering
    
    \begin{subfigure}[t]{0.45\textwidth}
        \centering
        \includegraphics[width=\textwidth]{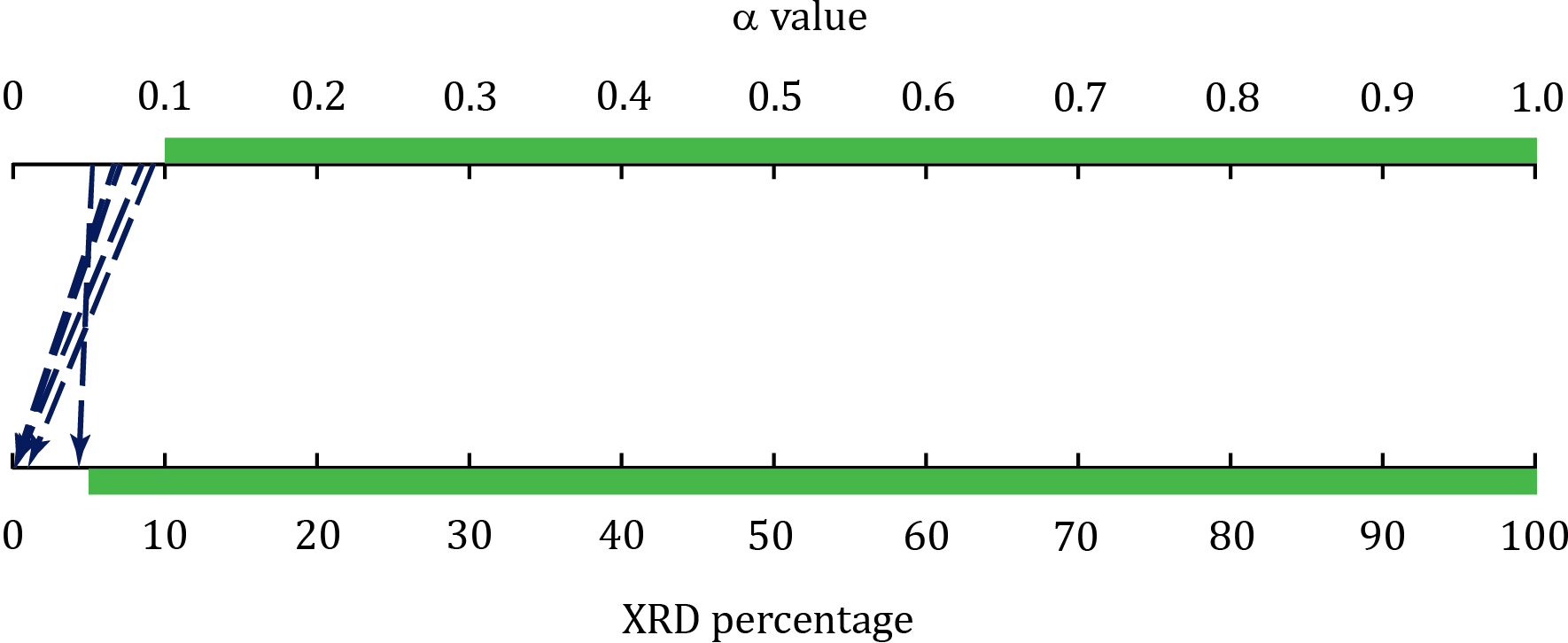}
        \caption{}
        \label{figure: map result site-1}
    \end{subfigure}
    ~
    \begin{subfigure}[t]{0.45\textwidth}
        \centering
        \includegraphics[width=\textwidth]{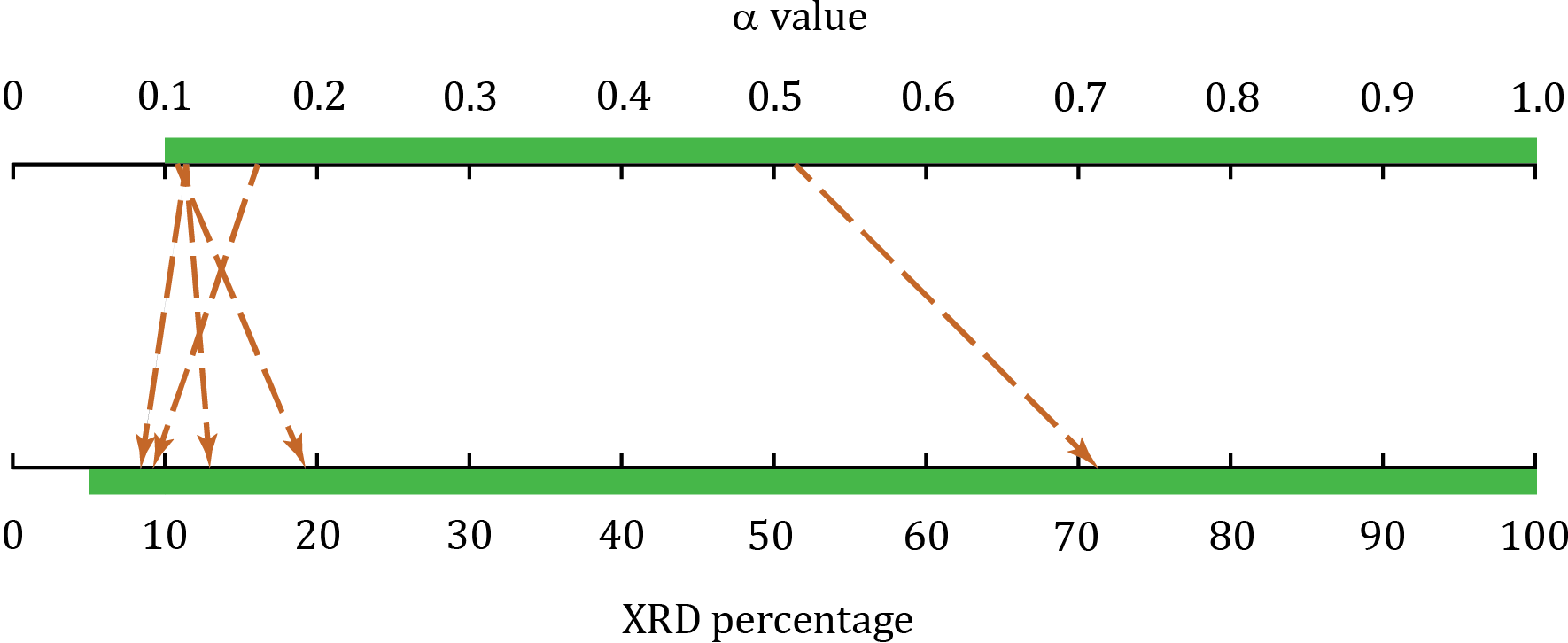}
        \caption{}
        \label{figure: map result site-2}
    \end{subfigure}
    
    \medskip
    
    \begin{subfigure}[t]{0.45\textwidth}
        \centering
        \includegraphics[width=\textwidth]{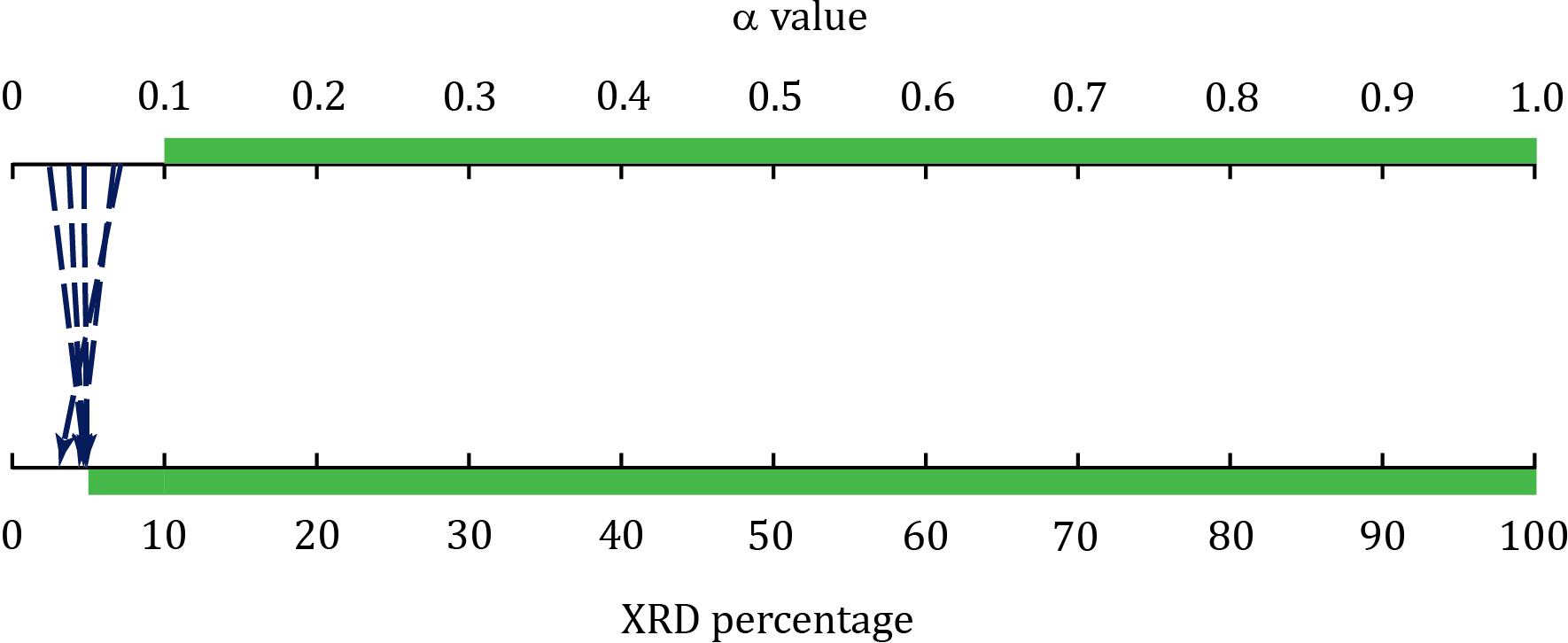}
        \caption{}
        \label{figure: map result site-3}
    \end{subfigure}
    ~
    \begin{subfigure}[t]{0.45\textwidth}
        \centering
        \includegraphics[width=\textwidth]{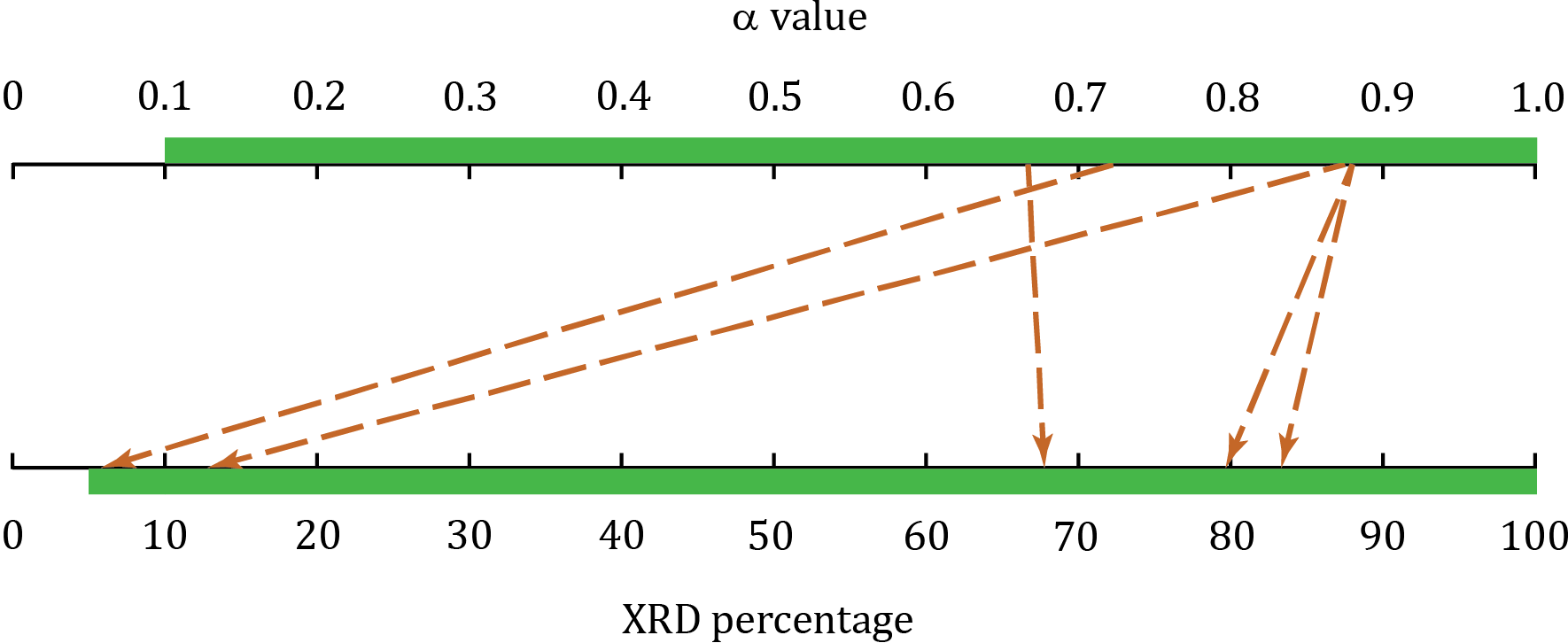}
        \caption{}
        \label{figure: map result site-4}
    \end{subfigure}
    
    \medskip
    
    \begin{subfigure}[t]{0.45\textwidth}
        \centering
        \includegraphics[width=\textwidth]{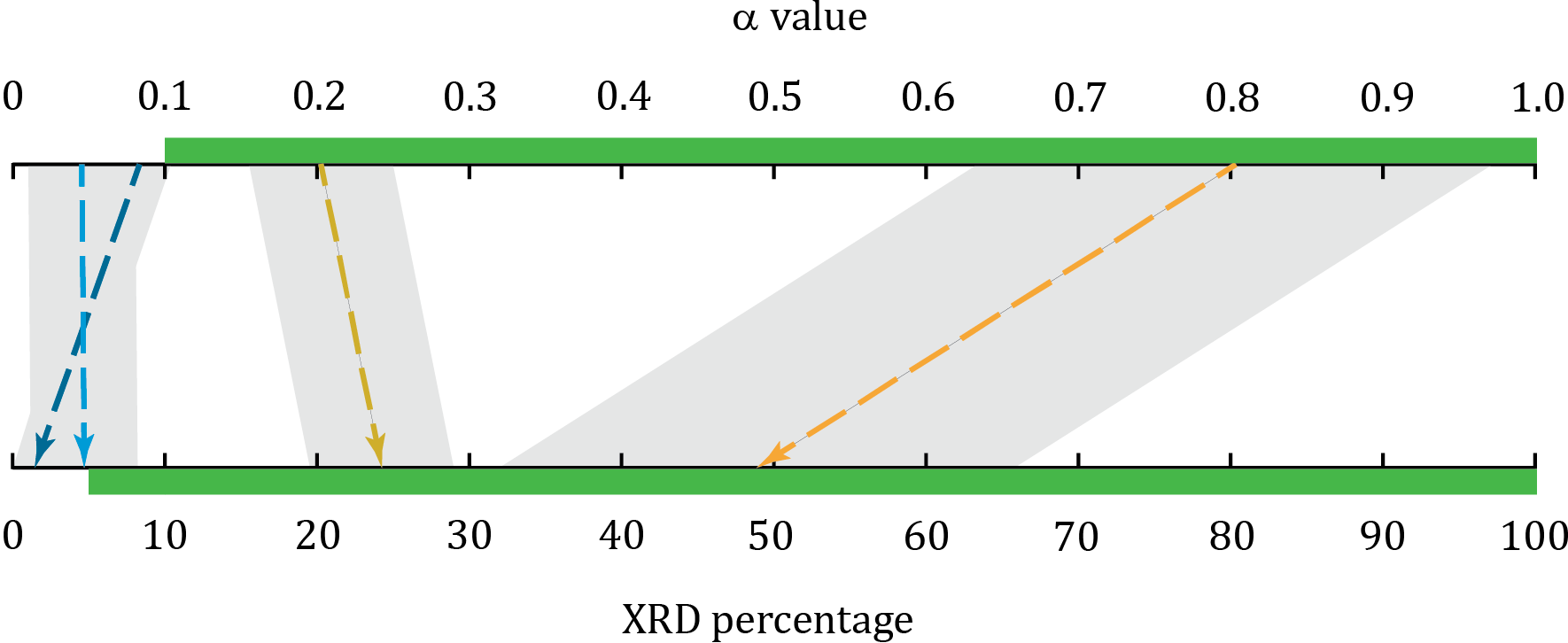}
        \caption{}
        \label{figure: map result site-overall}
    \end{subfigure}
    
    \medskip
    
    \begin{subfigure}[t]{0.9\textwidth}
        \centering
        \includegraphics[width=\textwidth]{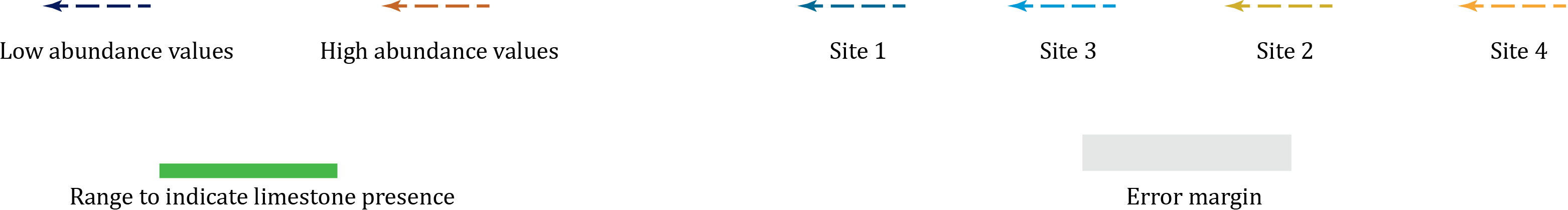}
    \end{subfigure}
    \caption{XRD results and $\alpha$ values of the soil samples of (a) site-1, (b) site-2, (c) site-3, (d) site-4, and (e) overall}
    \label{figure: xrd alpha mapping}
\end{figure*}

It was clear that the proposed method is capable of finding locations to survey and avoid in search of a particular mineral. However, to buttress the observations from these sites, the amount of limestone available according to the XRD test are provided in Table \ref{table: mean limestone parameter values}. Since, an objective of the work was to find locations with high and low probability for limestone, a qualitative classification of the pixels based on percentage weight thresholding is sufficient. With the mineral presence threshold of 5\,\% as suggested by \cite{mulder2013characterizing} and \cite{fantauzzi2011arsenopyrite}, if the XRD percentage of the sample and the $\alpha$ value of the corresponding pixel is higher than 5\,\% and 0.1, respectively; then it is assumed to contain limestone. The relation between the $\alpha$ values and the XRD results is presented in Fig. \ref{figure: xrd alpha mapping} for all four sites. According to Fig. \ref{figure: xrd alpha mapping}, it is discernible that the samples from the sites recognized as high probability for limestone by the algorithm have recorded higher XRD percentages than the weight threshold, conversely, XRD percentages lower than the threshold were recorded for samples with low probabilities for limestone. The premise that it is possible to subsume limestone if the XRD percentage of the sample and the $\alpha$ value of the corresponding pixel is higher than 5\,\% and 0.1, respectively, was found to be 100\% accurate from the conducted field test validation. Thereby confirming the proposed method's use as a precursor in decision making for site surveying. Though the proposed method has had better performances under the weight thresholding classification, it is customary to compare the results from the algorithm with the XRD results. In this regard, it is apt to calculate the correlation between the results as both methods calculate the probability to find limestone, but with disparate principles. The $\alpha$ values and the laboratory results had a correlation coefficient of 0.72 and a correlation of this magnitude is considered to be moderately strong.

Furthermore, the proposed algorithm underscored the extraction of a site-specific signature rather than a laboratory reference signature for limestone to calculate the abundances. To review the improvement from the suggested method, the accuracy with the XRD results was considered for the abundance values with and without the site-specific limestone reference extraction. The abundance values with the signal extraction recorded an accuracy of 0.757 while the other method had an accuracy of 0.458 along with a correlation coefficient of 0.31. The abundance map generated without the signal extraction is given in Fig. \ref{figure: mineral map without noise removal} in comparison with the mineral map produced from the proposed method. According to Fig. \ref{figure: mineral and contour maps} and Table \ref{table: mean limestone parameter values}, it is observable that without the stochastic cancellation-based signal extraction, higher $\alpha$ values have been computed for the pixels from the sites with experimentally low possibility for limestone.

\subsection{Application of the proposed work}
\label{subsection: application of the work}

In geology, to construct the terrain and lithological maps the general practice is to conduct walk-over surveys and perform laboratory tests \citep{hathaway1979us} on the collected samples. Even so, this practice is not the ideal way because the surveyors choose landmarks contingent on the professional experience to expedite the laborious and arduous task at hand. Next, once the samples are collected, their characteristics are examined and gleaned in a laboratory, thereafter georeferenced to the respective site. As the presence or absence of the mineral is decided after this cascade of procedures, it is possible to waste resources on sites with very little presence of the target mineral. In regard of that, the proposed method could be used to select locations to dodge when mineral presence is less likely as per the proposed method. Next, using the experimental results from the survey, the professionals could construct the contour plot for the lithology for separate minerals before generating the compounded survey map. Though these contour plots are generated from actual field data, it is not guaranteed that the entire contour will have the specified characteristics as these lines are extrapolated. In this regard, the proposed work in this paper derives the contour map for the limestone mineral using \remotesensing~ images through single-target identification approach similar to the first stage of the field surveys, and the presented algorithm is scalable and adaptable for other minerals too. Furthermore, the proposed method has attenuated possible overestimation of abundance of the target mineral when the stochastic cancellation and residual impurity removal is not performed. In addition to this, \remotesensing~ allows the user to scan through an entire region, thus the generated contour maps are consistent with the pixel characteristics, and nor are they extrapolated like in the field surveys. Once the algorithm is applied on a given \spectralimage{H}, it will generate the contour map for the survey site as well as the abundance map for the selected mineral. From the contour map and the abundance map, the surveyors can perform a more rigorous manual survey for further probing on locations that have $\alpha$ values larger than 0.1 as it is equivalent to high likelihood of mineral presence since \remotesensing~ data gives mostly surface mineral presence information. Besides, using the information from the laboratory experiments performed on the samples, the characteristics of the locations that was not surveyed can be estimated using regression techniques as the proposed method could produce a smooth abundance map as observed in Fig. \ref{figure: generated contour map}.

\section{Conclusion}
\label{section: conclusion}

This paper was focused on developing a methodology for mineral-indication with limestone as the case study using \remotesensing~data of the Jaffna peninsula of Sri Lanka. The work was aimed at single-target detection of the mineral in the presence of other mineral and non-mineral constituents. Furthermore, the proposed method does not require any prior knowledge regarding the composition of the survey area except for a laboratory reference signature of the target mineral. In the light of the adopted approach, a method to align a given set of soil pixels along the mineral purity increasing direction resulted from an intermediate step of the algorithm. The eigen-direction for purity alignment extends from the least correlated signatures to the most correlated signatures with the generic signature of the target mineral. Next, the proposed method derived a site-specific endmember for the principal mineral: limestone from the satellite image data using a Wiener filter arrangement and the laboratory-generated limestone signature. The site-specific endmember was used to compute the abundance values of limestone to generate a mineral map which was used to select site locations to be surveyed during a field survey. From the XRD test results of the collected samples from this field survey, it was observed that a considerable amount of limestone was available in sites that were given higher abundance values and vice versa, thus validated the usefulness of the proposed algorithms in mineral presence likelihood estimation. Importantly, the algorithm identified the likelihood of limestone from a site that was not surveyed for limestone from previous lithological surveys, and the derived results from the mineral map were supported by the XRD results from the field survey. The XRD threshold of 5\,\% coincided with the abundance threshold of 0.1 for mineral presence as validated by the results of the field survey.  The classification results for limestone presence and absence of the sites with the aforementioned thresholding yielded an accuracy of 100\% as there were no false presence or false absence indications of limestone by the proposed algorithm. This validated the use of the algorithm in the selection of sites to be surveyed for more likely sites as deemed by high $\alpha$ values (more than 0.1) and avoid for less likely sites as deemed by low $\alpha$ values (less than 0.1) before conducting the field survey. The algorithm is capable of estimating the abundance values despite the presence of other minerals as the values generated for the survey sites with the site-specific limestone signature recorded an accuracy of 75.7\% with the XRD results, whereas the abundances calculated without stochastic cancellation had an accuracy of only 45.8\%. Besides, the proposed classification method for low SNR \spectralimage{H}s prevailed over the application of \spectralimage{M}s and the inclusive nature of the single-target identification approach. Finally, the contributions of the proposed work can be enumerated as follows: 
(a) method to generate a mineral signature endemic to the survey site with stochastic cancellation and residual impurity removal.
(b) alignment of soil pixels according to the \relativemeasure~level of the target mineral.
(c) predictive modeling with the site-specific limestone signature as guidance for survey location selection with the $\alpha$ value criterion as high values are equivalent to more than 5\,\% of limestone presence.
(d) generation of a smooth contour map to be used to extrapolate mineral abundances of locations that were not surveyed.
(e) mitigation of possible overestimation or poor accuracy in \dlm~ from using a generic laboratory signature.
(f) single-target mineral detection of limestone in the presence of different minerals. 
Further, the proposed methodology can be extended to other minerals which will facilitate the multi-target detection of minerals as a cascade of single-target detection because the method development did not include assumptions specific to limestone. Also, the proposed algorithm is applicable to other uninterrupted geographical sites too. Nonetheless, additional research and field data are required to analyze the algorithm's performance with other minerals and geographical sites.

\section*{Authorship contribution statement}

\authorcont{yr} Methodology, Experiments, Writing - original draft.
\authorcont{hw} Experiments, Writing - original draft.
\authorcont{sh} Experiments, Writing - original draft.
\authorcont{vh} Conceptualization, Supervision, Writing - review \& editing.
\authorcont{rg} Conceptualization, Supervision, Writing - review \& editing.
\authorcont{pe} Conceptualization, Supervision, Writing - review \& editing.
\authorcont{as} Experiments, Supervision, Writing - review \& editing.
\authorcont{ly} Data curating, Supervision, Writing - review \& editing

\section*{Acknowledgment}
The authors acknowledge the support from the USGS for providing hyperspectral images from the EO-1 satellite's Hyperion sensor, and documents for standard preprocessing of the images. We are grateful to the Department of Geology, University of Peradeniya for arranging and providing equipment for soil sample preparation. The authors acknowledge the support received from Mr. Nevil Attanayake and Mr. Chanaka Perera during the fieldwork; and Mr. Kamal Jayasinghe of the Postgraduate Institute of Science, University of Peradeniya for conducting the laboratory tests.

\normalem
\small
\bibliographystyle{apalike}
\bibliography{bibliography}

\end{document}